\documentclass[fleqn,usenatbib]{mnras}
\usepackage{graphicx}	% Including figure files
\usepackage{amsmath}	% Advanced maths commands
\usepackage{amssymb}	% Extra maths symbols

\title[Pop III binary formation]
{Forming Pop III binaries in self-gravitating disks: \\
how to keep the orbital angular momentum}

\author[S. Chon et al.]{
Sunmyon Chon,$^{1}$\thanks{E-mail: sunmyon.chon@astr.tohoku.ac.jp}
and Takashi Hosokawa,$^{2}$
\\
$^{1}$Astronomical Institute, Tohoku University, Aoba, Sendai 980-8578, Japan\\
$^{2}$Department of Physics, Kyoto University, Kyoto 606-8502, Japan
}

\date{Accepted XXX. Received YYY; in original form ZZZ}

\pubyear{2019}

\begin{document}
\label{firstpage}
\pagerange{\pageref{firstpage}--\pageref{lastpage}}
\maketitle

\graphicspath{{./}}

\begin{abstract}
The disk fragmentation is a possible process leading to the formation of Population III stellar binary systems. However, numerical simulations show diverse fates of the fragments; some evolve into stable binaries and others merge away with a central star. To clarify the physics behind such diversity, we perform a series of three dimensional hydrodynamics simulations in a controlled manner. We insert a point particle mimicking a fragment in a self-gravitating disk, where the initial mass and position are free parameters, and follow the orbital evolution for several tens of orbits. The results show great diversity even with such simple experiments. 
Some particles shortly merge away after migrating inward, but others survive as the migration stalls with the gap-opening in the disk. 
We find that our results are well interpreted postulating that the orbital angular momentum is extracted by (i) the gravitational torque from the disk spiral structure, and (ii) tidal disruption of a gravitationally-bound envelope around the particle. Our analytic evaluations show the processes (i) and (ii) are effective in an outer and inner part of the disk respectively. There is a window of the gap-opening in the middle, if the envelope mass is sufficiently large. These all agree with our numerical results. We further show that the binaries, which appear for the ``survival'' cases, gradually expand while accreting the disk gas.
Our theoretical framework is freely scalable to be applied for the present-day star and planet formation.
\end{abstract}


\section{Introduction}
\label{sec_intro}

Massive black hole (BH) binaries recently discovered by gravitational wave (GW)
\citep[e.g.,][]{GW150914,GW151226,GW170104,GW170608,GW170814} may originate from massive stellar binaries born in the early universe. It has been suggested that progenitor stars should have had low metallicities $Z < 0.1~Z_{\odot}$, with which only the weak mass loss is expected \citep[e.g.,][]{Belczynski+2016}. Even metal-free, or Population~III (Pop~III), stars are one of possible candidates of their progenitors \citep[e.g.,][]{Belczynski+2004,Kinugawa+2014}. 

%--------------------------------------------------------------------------%

Theoretical studies predict that the formation of massive ($\sim 10-100~M_\odot$) stellar binaries should occur in the early universe, being initiated by gravitational fragmentation of the star-forming gas. 
An accretion disk growing around a protostar is thought to be a promising site of such fragmentation. In the case of Pop~III star formation, where H$_2$ molecules are the primary coolants, the disk fragmentation has been broadly observed in three-dimensional (3D) hydrodynamics simulations performed by different authors 
\citep[e.g.,][]{Saigo+2004, Machida+2008,Clark+2011,Susa2013,Hosokawa+2016,Riaz+2018,Susa2019}.
However, it also turns out that the orbital evolution of the fragments is very stochastic; some quickly migrate inward through the disk and merge away with the central star, while others survive for a long term to evolve into the secondary stars. 
\cite{Greif+2012} demonstrate that about one third of the fragments survive and others merge away, performing simulations with a very high spatial resolution \citep[see also][]{Stacy+2016}. Regardless of these developments, key physical processes behind the diverse evolution are yet to be understood. 
Unfortunately, the results are quantitatively very different among the authors.  For instance, the stellar multiplicity, i.e., the number of stars that survive for a while after the disk fragmentation, ranges from several \citep[e.g.,][]{Susa+2014} to almost a hundred \citep[e.g.,][]{Stacy+2016}.

%---------------------------------------------------------------------------%

The so-called Direct Collapse (DC) model, where the H atoms are the primary coolant, has been intensively studied as a possible star formation channel \citep[e.g.,][]{BL2003,Wise+2008,Regan+2014}.
Whereas the gas thermal evolution is different from the normal Pop III cases 
\citep[e.g.,][]{Omukai2001, Inayoshi+2014},
the disk fragmentation is also expected for such DC cases \citep{Matsukoba+2019}.
Indeed, recent 3D simulations demonstrate that the disk fragmentation does occur, showing the diverse fates of the fragments \citep{Becerra+2015, Sakurai+2016,Ardaneh+2018,Chon+2018,Suazo+2019}.
Some merge away with stars, but others survive in binary systems. In the DC cases, mean accretion rates onto protostars are $\sim 0.1$--$1~M_\odot~{\rm yr}^{-1}$.  
The protostars accrete $\sim 10^{4}$--$10^{5}~M_{\odot}$ of the gas in their lifetimes \citep[e.g.,][]{Latif+2015}, and finally collapse into intermediate-mass BHs (IMBHs) \citep[e.g.,][]{Shibata+2002,Uchida+2017}. Very massive stellar binaries, if any, may evolve into IMBH binaries. They are being targeted by future space GW facilities such as LISA \citep{Amaro-Seoane+2012} and DECIGO \citep{Kawamura+2011}.

%--------------------------------------------------------------------------%

Limitations in numerical methods hinder one from clarifying the physics behind the evolution after the disk fragmentation.
The sink particle method \citep[e.g.][]{Bate+2002}, which has been often used to follow a long-term evolution of the protostellar accretion, is not an exception. 
In this method, a protostar and its very vicinity are masked by a sink particle with a finite size. The computational cost is substantially reduced without spatially resolving the dense gas. As a trade-off, however, an insufficient spatial resolution prevents one from evaluating the tidal torque acting on a dense part of the gas. The migration of the fragments, a key process to control the evolution after the disk fragmentation, should be affected by such a limitation \citep{Greif+2012}.
Moreover, results should be intrinsically resolution-dependent, i.e., depend on below which threshold density the gas is actually resolved. To make matters worse, there are different criteria on the creation and merger of sink particles. Such uncertainties may lead to a different resolution-dependence of the results.
It is thus difficult to extract true physical processes that may cause the divergent evolution.  

%----------------------------------------------------------------------%

In this paper, we aim to clarify physical processes that control the divergent evolution, disentangling the issues described above. 
On this purpose, we focus on the evolution of the circumstellar disk just after the disk fragmentation. The fragments are approximately represented by point particles, whose initial masses and positions are freely parameterized. In such a well organized manner, we systematically follow the orbital evolution of each particle for several tens of orbits performing a suite of hydrodynamics simulations.
We do not adopt the sink particle method to avoid numerical uncertainties that are not fully under control. We instead assume a stiff equation of state (EOS) above a threshold density \citep[e.g.,][]{Hirano+2017}, supposing the DC cases. The inserted particle accretes the gas as it migrates through the disk, and develops a surrounding envelope bound by its gravity. We spatially resolve the interior structure of such a bound object (called ``clump''). 
Even with the above simple numerical experiments, our results show the divergent orbital evolution as reported in the literature; some clumps merge away after the inward migration, and others survive for a while joining binary systems. 

%-----------------------------------------------------------------%

The rest of the paper is organized as follows. In Section~\ref{sec_methodology}, we describe our methodology. Our numerical results are presented in Section~\ref{sec_results}. We show that the divergent evolution observed in our simulations are actually well interpreted by considering physical processes that can remove the orbital angular momentum of the migrating clumps. We present thorough comparisons between the numerical results and such analytic considerations in Sections~\ref{sec_qa_diagram} and \ref{sec_final_separation}.
In Section~\ref{sec_discussion}, we discuss limitations and implications of the current work.
We finally summarize our findings in Section~\ref{sec_conclusion}.

%------------------------------------------------------------------%

%%%%%%%%%%%%%%%%%%%%%%%%%%
%%%%%%%%%%%%%%%%%%%%%%%%%
\section{Methodology} 
\label{sec_methodology}
%%%%%%%%%%%%%%%%%%%%%%%%%%
%%%%%%%%%%%%%%%%%%%%%%%%%%

We perform a suite of 3D hydrodynamics simulations using the N-body + Smooth Particle Hydrodynamics code, {\tt Gadget2} \citep{Springel2005}. We assume the following barotropic description of the equation of state (EOS)
\begin{align} \label{eq_EOS}
P \propto \left \{ 
\begin{array}{ll}
\rho          \;\;\; & (\rho < \rho_\text{adib}), \\
\rho^{5/3} \;\;\; & (\rho \geq \rho_\text{adib}),
\end{array}
\right .
\end{align}
where $P$ and $\rho$ are the pressure and density of the gas, and $\rho_\text{adib}$ is the critical density, above which the gas behaves adiabatically. We assume the constant temperature $T_{0} = 8000~$K for $\rho < \rho_\text{adib}$ and set $\rho_\text{adib} = 2 \times 10^{-8}~\mathrm{g~cm}^{-3}$, which mimics the thermal evolution of a dynamically collapsing cloud for the DC cases \citep[e.g.,][]{Omukai2001, Inayoshi+2014}. The critical density corresponds to the number density $n_\text{adib} = 10^{16}~\mathrm{cm^{-3}}$.
The above prescription ignores the radiative feedback from accreting protostars, which potentially heats up the gas. We only consider the early stage when the stellar mass is less than $10~M_\odot$ and the resulting radiative feedback is weak.

%-----------------------------------------------------------------------------%

Since the only dimensional constant relevant to our calculations is the gravitational constant $G$,
we can freely rescale our simulation results by varying $T_{0}$ and $n_\text{adib}$.
By such rescaling, we can apply our results to other cases where the disk fragmentation occurs, e.g., the normal Pop~III and even present-day star formation. We discuss such applications later in Section~\ref{sec_scale}.

\begin{figure*}
	\centering
		\includegraphics[width=18.cm]{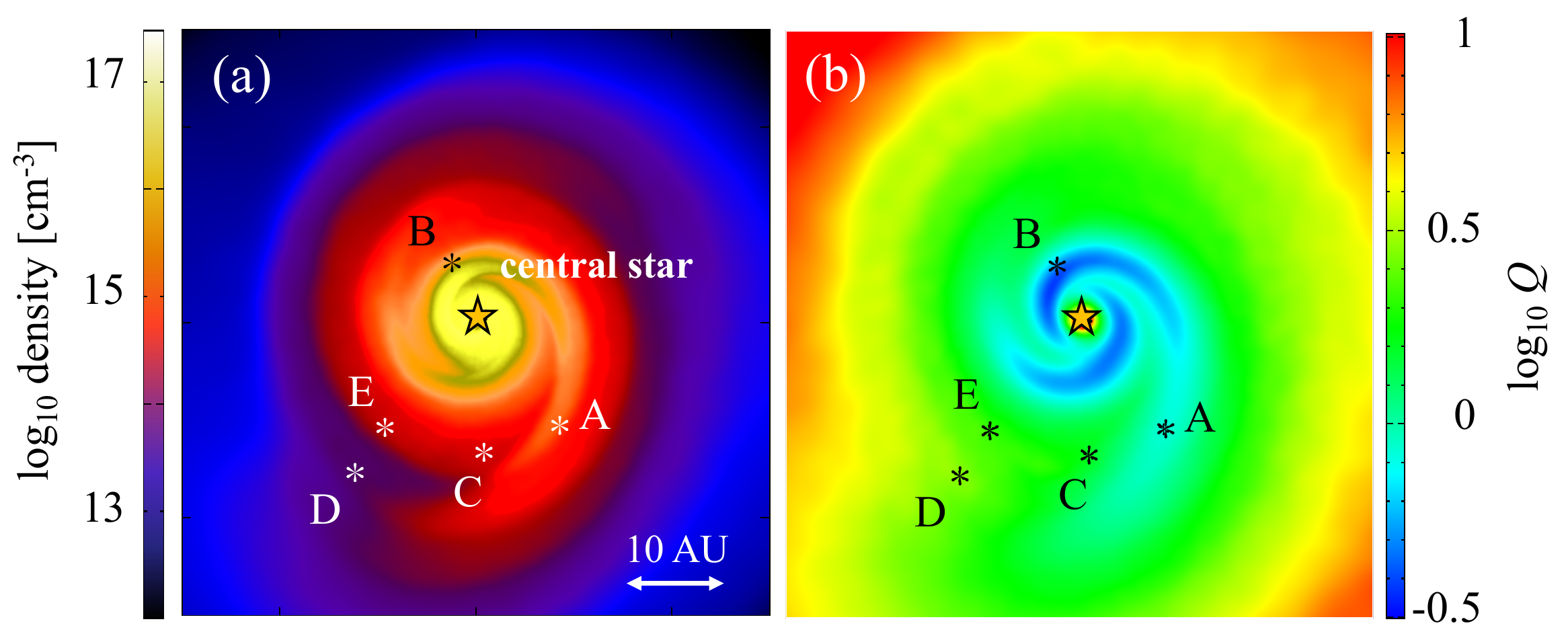}
		\caption{The spatial distributions of (a) the gas density in the disk midplane and (b) the Toomre-$Q$ parameter (eq.~\ref{eq:toomreq}) at the epoch of $4~$years after the birth of the central star, when we insert the point particle whose subsequent orbital evolution is followed. The initial positions of the point particles are represented by the asterisks, which are labeled as A -- E.}
		\label{fig_initial_condition}
\end{figure*}

\subsection{Set-up of a star-disk system}
\label{ssec_setup}

Our aim is to investigate the orbital evolution of fragments in a self-gravitating disk, supposing the evolution after the disk fragmentation. To this end, we first set up a star-disk system by also performing a preparatory simulation. We first follow the dynamical collapse of a gas cloud, which leads to the birth of a protostar. We then insert a sink particle to mask a central part including the protostar, and further follow the growth of a circumstellar disk for a while. We separately describe each of such sequential simulations below. 

\subsubsection{Early collapse stage of the gas cloud} 
\label{sssec_collapse}

We start our simulation with an unstable Bonnor-Ebert sphere with a rigid rotation. The initial cloud is characterized by the two parameters, $\alpha=E_\text{therm}/E_\text{grav}$ and $\beta=E_\text{rot}/E_\text{grav}$, the ratios of the thermal ($E_\text{therm}$) and the rotational energy ($E_\text{rot}$) to the gravitational energy ($E_\text{grav}$), respectively.
We consider a specific case with $\alpha=0.6$ and $\beta=0.2$. Such a cloud represents those found by \citet{Chon+2018}, who have simulated the formation of the DC clouds, where the H atomic cooling causes the collapse, in a full cosmological context.

%------------------------------------------------------------------------------%

We set the initial central density of the cloud as $n_{0}= 10^{11}~\mathrm{cm}^{-3}$, with which the total cloud mass and radius are $M_0 \simeq 57~M_{\odot}$ and $R_0 \simeq 220~$AU, respectively.
Note that the above cloud mass is much smaller than that of the typical DC clouds, $\sim 10^{5}~M_{\odot}$ \citep[e.g.,][]{Chon+2016}. Nonetheless, our choice of the cloud parameters are reasonable for the current work, where we only consider an earliest epoch of the protostellar accretion stage. We return this point in Section~\ref{ssec_particle}.

%------------------------------------------------------------------------------%

Soon after we begin the calculation, the cloud collapses in a so-called self-similar ``run-away'' fashion \citep{Larson1969}. The density at the cloud center significantly rises throughout this stage. When the central density reaches $n_\text{adib} = 10^{16}~\mathrm{cm}^{-3}$, 
we insert a sink particle with the radius of $R_\text{sink}=0.5~$AU at the density maxima, supposing the birth of the protostar. We refer this particle as the ``central star'' hereafter.

\subsubsection{Later protostellar accretion stage} 
\label{sssec_accretion}

After inserting the central star particle, its mass begins to grow via accretion. Since the accreting gas has a finite angular momentum, a rotationally supported disk appears around the central star.
Fig.~\ref{fig_initial_condition} (a) shows the density distribution around the central star at $t=4~$years, where the time origin corresponds to the epoch of the birth of the protostar.
The masses of the central star and the surrounding disk are 
$5.7$ and $10~M_{\odot}$, respectively.
The time-averaged mass accretion rate onto the central star is $\sim 0.1~M_\odot~{\rm yr}^{-1}$, which is the typical value expected in the DC model.

%--------------------------------------------------------------------------%

The disk is marginally stable against the gravitational instability.
To see this, we evaluate the Toomre-$Q$ parameter,
\begin{align}
Q = \frac{c_\text{s} \Omega}{\pi G \Sigma},
\label{eq:toomreq}
\end{align}
where $\Omega$ is the orbital frequency and $\Sigma$ is the surface mass density. Fig.~\ref{fig_initial_condition} (b) shows the distribution of the Toomre-$Q$ value.
The Toomre-$Q$ value averaged over the entire disk is close to the unity, which indicates the disk is self-gravitating. The disk has the non-axisymmetric spiral structure, along which the Toomre-$Q$ is much smaller than the unity. However, the disk self-gravity is not strong enough to cause the gravitational fragmentation.

%---------------------------------------------------------------------%

\begin{table}
\begin{center}
\begin{tabular}{lccl}
\hline
Model & initial position & mass [$M_{\odot}$]  & final fate\\
\hline
\\
M005A & A & 0.05 & merge \\
M005B & B & 0.05 & merge \\
M005C & C & 0.05 & merge \\
M005D & D & 0.05 & merge \\
M005E & E & 0.05 & merge \\
\hline \\
M01A & A & 0.1 & merge \\
M01B & B & 0.1 & merge \\
M01C & C & 0.1 & survive \\
M01D & D & 0.1 & survive \\
M01E & E & 0.1 & merge \\
\hline \\
M02A & A & 0.2 & survive \\
M02B & B & 0.2 & merge \\
M02C & C & 0.2 & survive \\
M02D & D & 0.2 & survive \\
M02E & E & 0.2 & survive \\
\hline
\end{tabular}
\caption{The properties of models considered in this paper.
}
\label{t:models}
\end{center}
\end{table}

\subsection{Inserting a point particle in the disk}
\label{ssec_particle}

We insert a point particle in the disk at the epoch presented in Fig.~\ref{fig_initial_condition}. The particle resembles small self-gravitating bodies generated by the disk fragmentation.
The epoch of inserting the particle is arbitrary, but the current case satisfies conditions (i) the Toomre-$Q$ value averaged over the disk is around the unity, and (ii) it is soon after the formation of the star-disk system.
We have considered point (i) because imposing $Q \sim 1$ well represents the time-averaged state of a self-gravitating disk \citep[e.g.][]{Lodato+2005}, even if the disk fragmentation temporarily occurs where $Q < 1$ \citep[e.g.,][]{Hosokawa+2016,Chon+2018}. Point (ii) comes from the fact that the disk fragmentation has been often observed for an early phase of the accretion stage \citep[e.g.][]{Susa+2014}. Protostellar radiative feedback effects, which are neglected in our work, should be minor in such an early phase when the stellar mass does not exceed $10~M_\odot$.

%------------------------------------------------------------------------%

The initial positions of the point particles are marked by the asterisks in Fig.~\ref{fig_initial_condition}. The positions A and B are located on a spiral arm, where $Q \lesssim 1$, considering the fact that the disk fragmentation typically occurs through the fragmentation of such spiral-arm structure \citep[e.g.][]{Sanemichi+2016,Inoue+2018}. 
We also study other cases with the positions C, D, and E, which are located near the edge of the Keplerian disk with $Q \sim 1$.
The radial separations between the central star and the positions A, C, and E are almost the same. The initial masses of the point particles are chosen as $M_\text{p} = 0.05$, $0.1$, and $0.2~M_{\odot}$, which are much smaller than the masses of the central star and the disk.
We consider cases with these different initial masses for each initial position A -- E.
All the examined cases are summarized in Table~\ref{t:models}.
The initial velocity of the particle is set to be the same as the local velocity field of the gas.

%--------------------------------------------------------------------------------%

As described later, our simulations show that the inserted particles initially migrate inward toward the central star in all the cases.
When the particle's distance from the central star becomes smaller than the resolution limit $\simeq 0.5~$AU, we stop a simulation run assuming the particle merges with the central star. 
We consider that the particle ``survives'' unless it merges away in $t_{0}=15~$years. 
In such a case, the migration stalls at some point and the pair of the particle and the central star forms a binary system.
The duration of $t_{0}=15~$years corresponds to about three times the orbital timescale of the initial disk. We note that, for all the models in which the particle experiences the merger, it occurs in the timescale comparable to the disk orbital time.

%--------------------------------------------------------------------------%

Recall that we have prepared the star-disk system starting with the small cloud with $M_0 \simeq 57~M_\odot$ (Section~\ref{sssec_collapse}). Such a small mass is sufficient to follow the evolution for $\lesssim 100$~years of the short earliest period in the protostellar accretion stage. Even with the current settings, in fact, the mass accretion onto the central star-disk system continues for $\simeq 300~$years, about the free-fall timescale of the initial cloud core.
In the realistic cases of the DC model, however, the mass accretion should continue for $\sim 10^6~$years 
after the dynamical collapse of massive clouds with $\sim 10^5~M_\odot$. Such long-term evolution of the surviving binaries is out of the scope of the current work, and is briefly discussed in Appendix~\ref{sec_revo_long}.

%-------------------------------------------------------------------------------%

\begin{figure}
		\includegraphics[width=8.2cm]{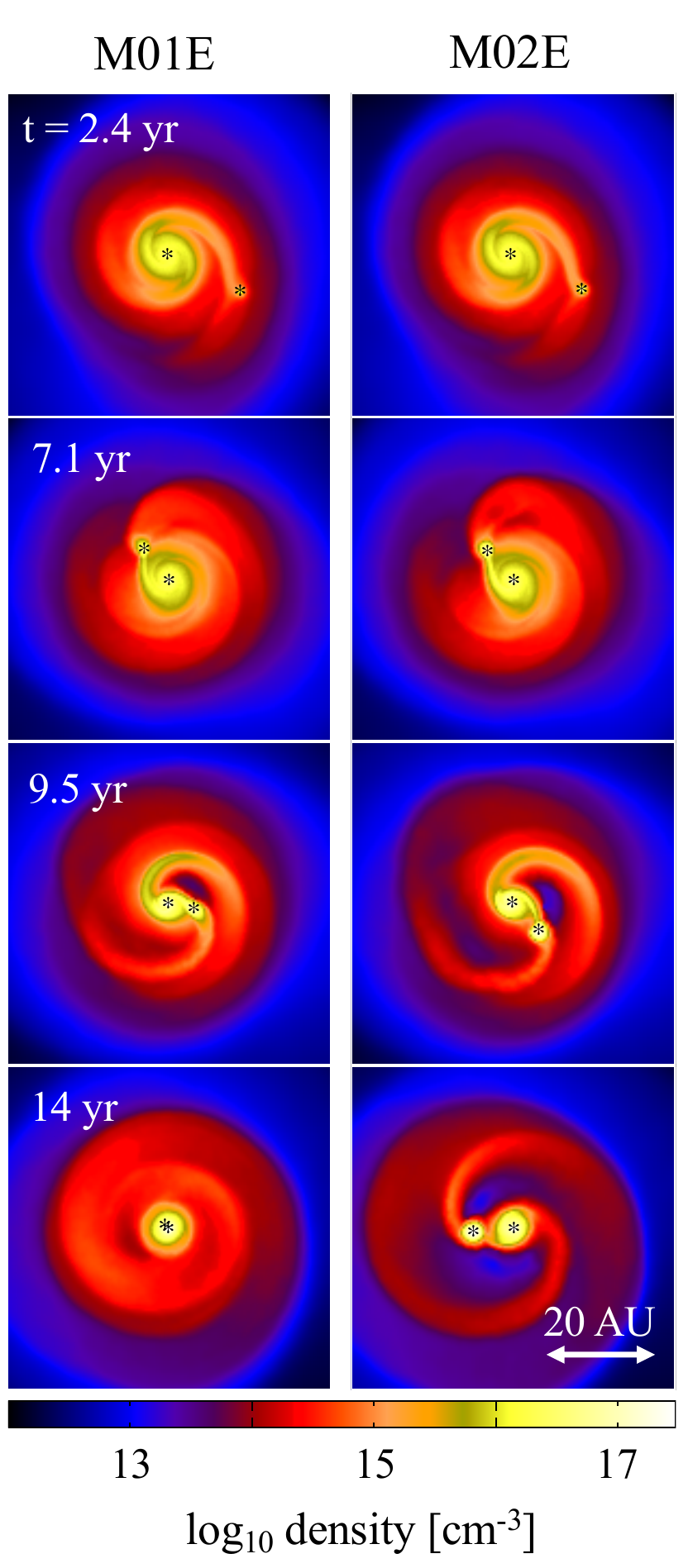}
		\caption{Time evolution of the density distributions around the central star after inserting a point particle with the mass of $0.1$ (left) and $0.2~M_{\odot}$ (right) at the position E (see Fig.~\ref{fig_initial_condition}), i.e., for the cases of M01E and M02E (Table 1).
The color scale represents the number density of the hydrogen atom in the disk midplane. The elapsed time after inserting the particle are presented in the left panels. M01E and M02E are the typical cases of the ``merger'' and ``survival'' of the inserted particles (see text). }
		\label{fig_snapshots}
\end{figure}

\begin{figure}
		\includegraphics[width=7.5cm]{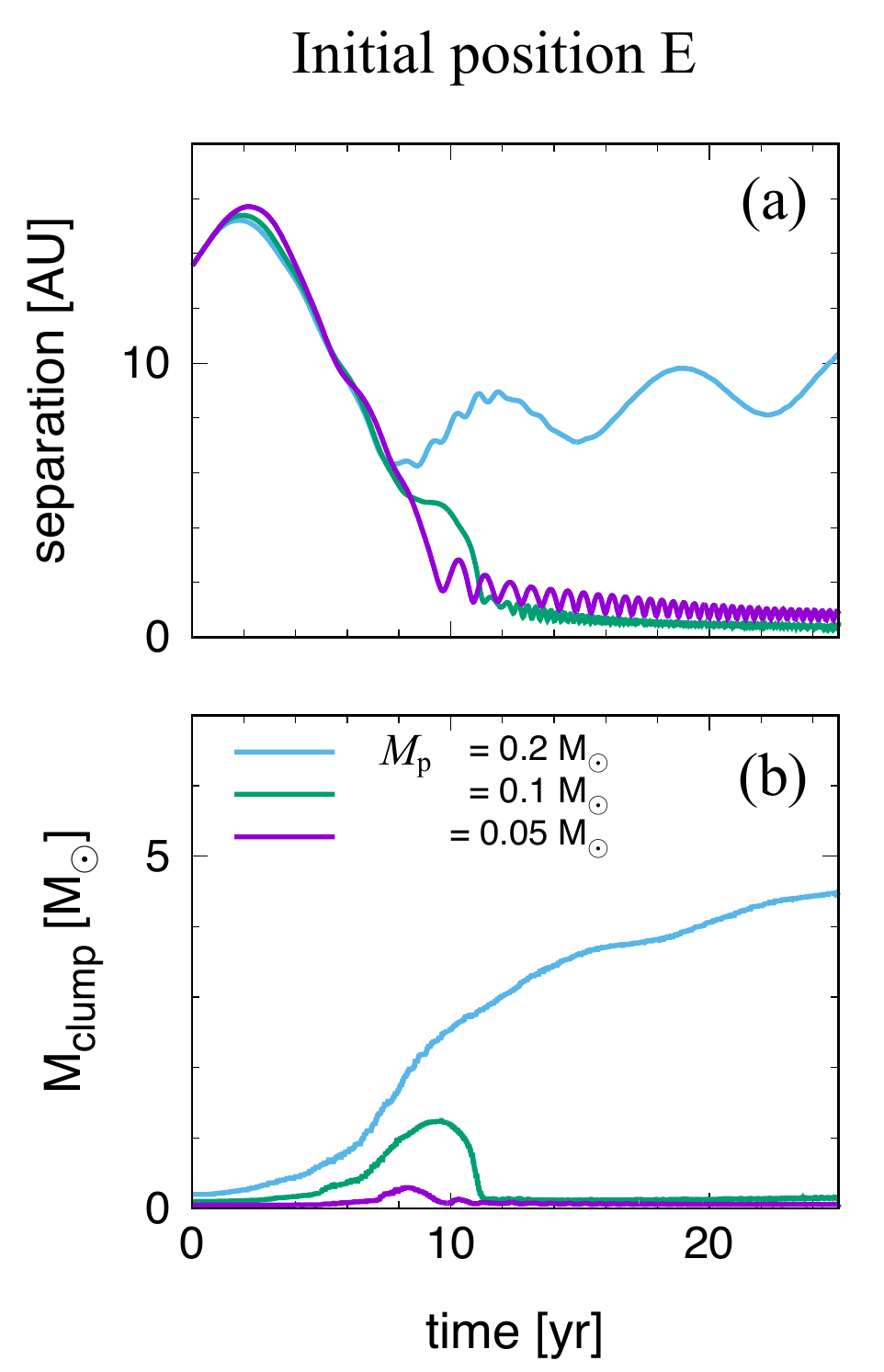}
		\caption{Time evolution of (a) the separation between the central star and the point particle and (b) the clump mass for models M005E (purple), M01E (green), and M02E (blue), where the initial mass of the particle is $0.05$, $0.1$, and $0.2~M_\odot$ and the initial position is E (Fig.~\ref{fig_initial_condition}). The clump mass is defined as the total mass of the point particle and the surrounding adiabatic envelope (also see text).}
		\label{fig_M_Revo}
\end{figure}

%%%%%%%%%%%%%%%%%%%%%%%
%%%%%%%%%%%%%%%%%%%%%%%
\section{Results} 
\label{sec_results}
%%%%%%%%%%%%%%%%%%%%%%%
%%%%%%%%%%%%%%%%%%%%%%%

\subsection{Migration of ``Clumps''}
\subsubsection{Visual Inspection}

We first focus on the orbital evolution of the particles with the different initial masses starting with the same initial position E. Fig.~\ref{fig_snapshots} shows the time evolution of the disk density structure for such cases of M01E (left) and M02E (right), where the initial masses are $M_\text{p} = 0.1$ and $0.2~M_{\odot}$. Fig.~\ref{fig_snapshots} does not show case M005E, where the initial mass is $0.05~M_{\odot}$, because the evolution is almost the same as that for case M01E.
Fig.~\ref{fig_M_Revo}~(a) presents the evolution of the separation between the particle and the central star. As seen in the figures, the orbital evolution of the particle for the initial $\lesssim 10$~years is very similar between these models; the particle migrates inward through the disk. 

%--------------------------------------------------------------------------------%

In contrast, the orbital evolution for $t \gtrsim 10~$years depends on the initial mass. 
When the stellar mass is $0.1~M_{\odot}$, the particle continuously migrates toward the central star.
The separation finally reaches our resolution limit, so that the particle merges with the central star. When the stellar mass is $0.2~M_{\odot}$, however, the inward migration stalls and the separation does not shrink any more. 
Moreover, the separation turns to increase afterwards as presented in Fig.~\ref{fig_M_Revo}~(a). 

%---------------------------------------------------------------------------------%

During the inward migration, a part of the disk gas becomes gravitationally bound by the point particle. As a result, the particle is enshrouded by the adiabatic envelope. 
We call this structure (the particle + gravitationally-bound gas) as ``migrating clump'' or simply ``clump'', the bound gas as ``envelope'', and the point particle as ``core'' of this clump.
Fig.~\ref{fig_M_Revo}~(b) presents the evolution of the mass of the migrating clump.
For $M_\text{p} = 0.2~M_{\odot}$ (case M02E), 
the clump continuously accretes the disk gas to have an envelope much more massive than the core. The migrating clump consequently opens up a large gap in the disk by the epoch of $t=9.5$~years, as shown in Fig.~\ref{fig_snapshots}.

\begin{figure}
	\centering
		\includegraphics[width=7.0 cm]{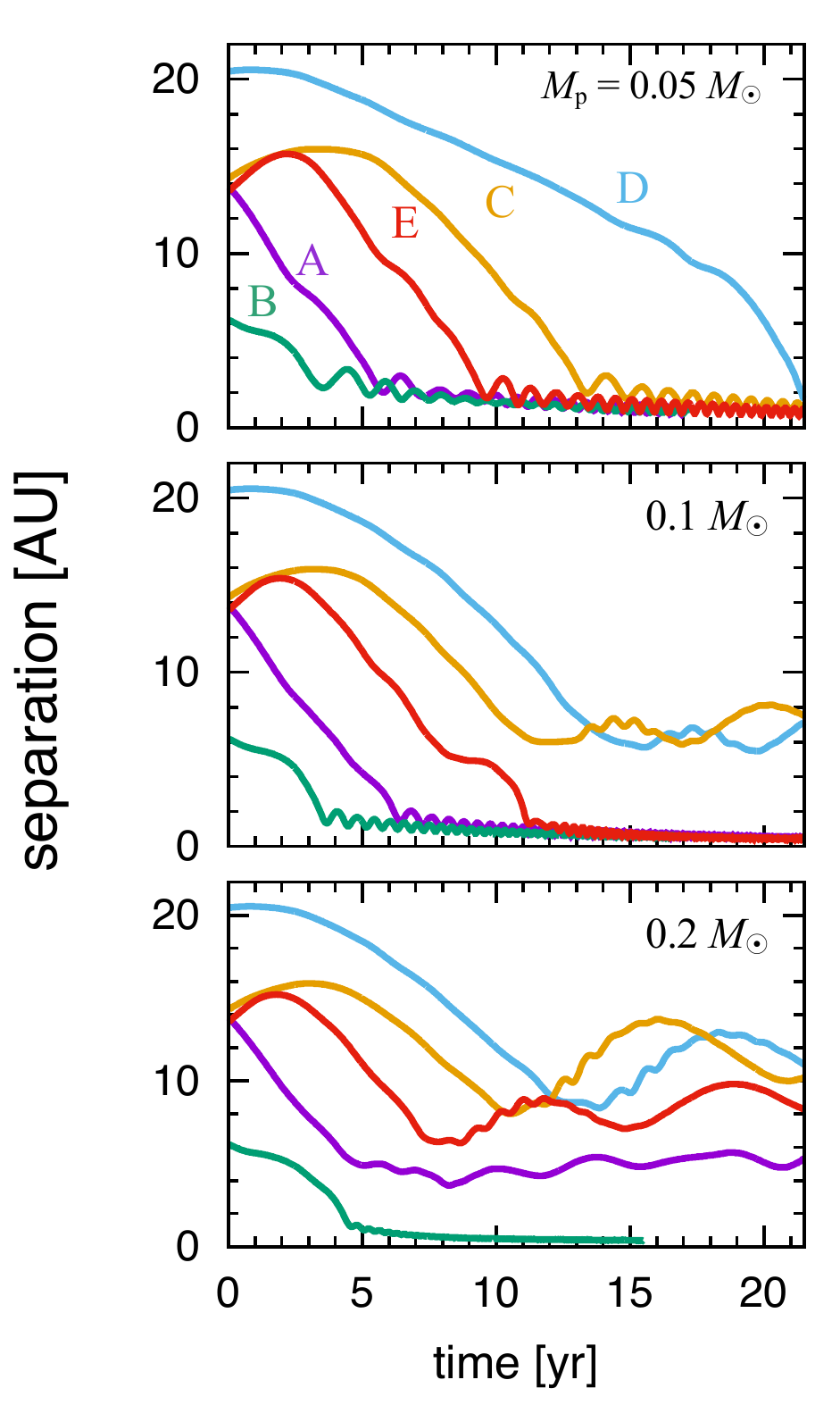}
		\caption{Variations of the orbital evolution of the inserted particles with the different initial masses and positions. The three panels present the time evolution of the separations between the central star and the particle for the different initial masses, $0.05$ (top), $0.1$ (middle), $0.2~M_\odot$ (bottom panel). In each panel, the different lines represent the different initial positions A (purple), B (green), C (yellow), D (blue), and E (red), as also remarked in Fig.~\ref{fig_initial_condition}. When the separation becomes smaller than $0.5~$AU, we stop the simulations assuming the clump merges with the central star.}
		\label{fig_sep_short}
\end{figure}

\begin{figure}
	\centering
		\includegraphics[width=7.5cm]{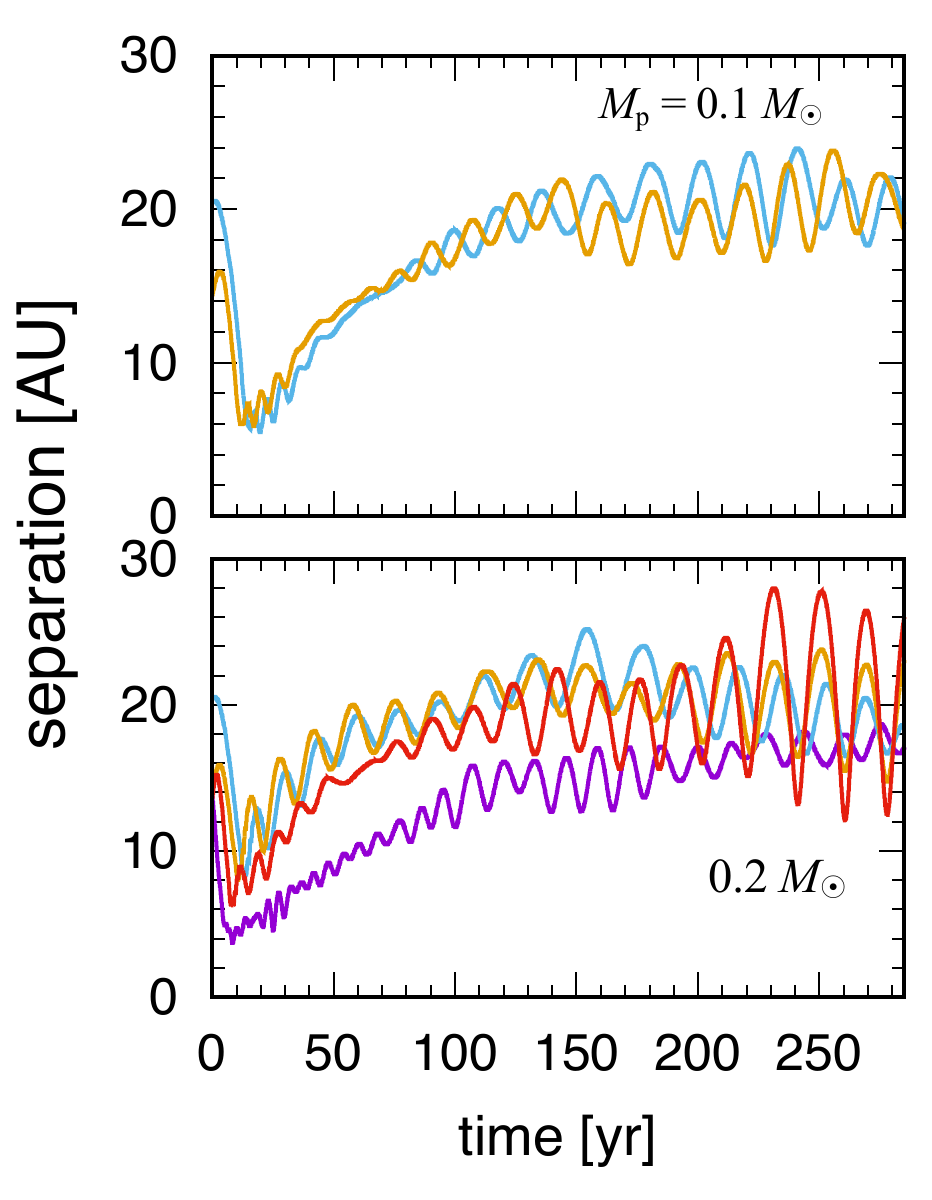}
		\caption{The long-term evolution of the separations between the central star and the inserted particles for the cases where the clumps ``survive'' without merging away. 
The top and bottom panels present the cases with the initial masses of $0.1$ and $0.2~M_\odot$. We do not show the cases with the smallest initial mass $0.05~M_\odot$ because the clumps merge away with the central star for such cases. In each panel, the different line colors represent the same different initial positions as in Fig.~\ref{fig_sep_short}.}
		\label{fig_sep_long}
\end{figure}

\subsubsection{Variations of the orbital evolution}

Fig.~\ref{fig_sep_short} shows the time evolution of the separation between the clump core and central star for all the examined cases.
Regardless of the simple setup of our simulations, the results show diversity of the clump orbital evolution. Nevertheless, there is still a trend; the clumps have more chances of the survival with the higher initial masses. 
For example, the clumps migrate to merge with the central star for the smallest initial mass $M_\text{p}=0.05~M_{\odot}$, while only one case ends up with the merger for the largest mass $0.2~M_\odot$ (case M02B).
Such a trend has been actually reported in studies on the migration process of planets and BHs through gas disks \citep[e.g.,][]{Escala+2005,Crida+2006,DelValle+2012}. 
They all show that the migration speed often drops when the massive objects create gaps by their strong gravitational perturbations. 
We have also confirmed such a gap-like structure for our all cases where the migration stalls. The snapshot at $t=9.5~$years in Fig.~\ref{fig_snapshots} presents an example. 

%---------------------------------------------------------------------------------%

Varying the initial position also changes the fates of the clumps.
With the same initial mass of $M_\text{p}=0.1~M_{\odot}$, for instance, the clump survives with the initial positions C and D while it merges away for the other cases (middle panel of Fig.~\ref{fig_sep_short}). 
Note that the initial separations between the central star and the positions A, C, and E only differ by 3\%. Nonetheless, the resulting evolution is largely different among them. 
The clump starting from position A, located along the spiral arm,
migrates inward the fastest, since the spiral arm extracts the angular momentum efficiently.
In the other cases, the clump are located apart from the spiral arms.
They first migrate outward and the inward migration begins just after they join the spiral arm.

%----------------------------------------------------------------------------------%

Although not shown in Fig.~\ref{fig_sep_short}, we have continued to follow the long-term evolution for cases where the clump does not merge with the central star. 
Fig.~\ref{fig_sep_long} shows such evolution of the separations between the central star and clump for $\simeq 300$~years, when the mass accretion ceases owing to our artificial set-up (see Section~\ref{sssec_collapse}). 
The figure shows that the inward migration stalls at $\simeq 5 - 10$~AU from the central star in the initial $\sim 10$~years. 
This timescale approximately corresponds to one orbital time, during which the gap-like structure develops around the massive clumps for the presented cases. 
Afterwards, the clump and central star are in a binary system. The binary separation turns to increase, and becomes almost constant of $10$--$20$~AU in $\sim 100$~years.
Such an outward migration has been also reported in previous studies on the star/planet formation 
\citep[e.g.,][]{Lin+2012, Zhu+2012, Nayakshin2017,Stamatellos+2018}. We investigate this phenomenon later in Section~\ref{sec_qa_diagram}.

%%%%%%%%%%%%%%%%%%%%%%%%%%%%%%%%%%%%%%%%%%%%%%%%
\subsection{Analytic Considerations: Physics of the Clump Migration} 
\label{sec_phys}
%%%%%%%%%%%%%%%%%%%%%%%%%%%%%%%%%%%%%%%%%%%%%%%%

In this section, we aim to clarify physical processes which cause the divergent evolution of the migrating clumps. To this end, we analytically consider how the orbital angular momentum of the clumps can be kept or extracted in self-gravitating disks.

\subsubsection{Migration and gap formation} 
\label{sec_gapformation}

In our simulations, the inserted clump initially migrates inward for all the examined cases. The disk has the spiral structure during that as presented in Fig.~\ref{fig_snapshots}. It seems that the gravitational torque owing to such a non-axisymmetric structure drives the inward migration. 
The similar phenomenon has been reported in previous studies in different contexts, including the present-day star/planet formation \citep[e.g.,][]{GT1980,Tanaka+2002,Paardekooper+2010,Baruteau+2011,Zhu+2012}, and the massive BH merger in galactic nuclei \citep[e.g.,][]{Escala+2004, Escala+2005}.
We refer the early migration as ``Type-I migration'' hereafter, the conventional term indicating the migration caused by a planet whose mass is too small to largely disturb the disk structure. Note that the Type-I migration normally supposes non-self-gravitating disks where the spiral structure is excited by the planet. In contrast, the similar structure is excited as a global mode of the disk gravitational instability in our cases. 
Nonetheless, the timescale of the early inward migration seen in our cases is $10~$years, which is comparable to that of the conventional Type-I migration \citep{Inayoshi+2014, Latif+2015b, Latif+2015c}.

%-----------------------------------------------------------------------------------------------------------------%
We have already shown that clearing the gap in the disk marks the end of the early inward migration (see case M02E in Fig.~\ref{fig_snapshots}). This is again similar to how Type-I migration ceases in proto-planetary disks \citep[e.g.,][]{Kanagawa+2018}. Since the planet mass rises with the mass accretion, the disk structure is strongly perturbed by its gravity in the end. The Type-I migration rate significantly drops after the gap opening for such cases. 
%-------------------------------------------------------------------------------------------------------------------%

One of the necessary conditions for the gap opening is called the thermal condition \citep{Lin&Papaloizou1993},
\begin{align}
H_\text{d} < R_\text{Hill},
\end{align}
where $H_\text{d}$ is the disk scale height and $R_\text{Hill}$ is the Hill radius of the migrating clump.
Assuming the disk with the Kepler rotation, we analytically evaluate the disk scale height as,
\begin{align} \label{eq_Hdisk}
H_\text{d} &= c_\text{s} \sqrt{\frac{a^{3}}{G M}} \nonumber \\
&= 3.3~\mathrm{AU} \left ( \frac{c_\text{s}}{10~\mathrm{km~s^{-1}}} \right ) \left ( \frac{a}{10~\mathrm{AU}} \right )^{3/2} \left ( \frac{M_{*}}{10~M_{\odot}} \right )^{-1/2},
\end{align}
where $a$ is the separation between the central star and the migrating clump and $M_*$ is the central stellar mass. The Hill radius is defined as the size of the region where the gravity from the clump dominates over the tidal force from the central star. It is well fitted by the following formula \citep{Eggleton1983},
\begin{align} \label{eq_rHill}
R_\text{Hill} &= a \frac{0.49 q ^{2/3}}{0.6 q^{2/3} + \log (1+q^{1/3})},
\end{align}
where $q$ is the mass ratio of the central star to the clump.

%-------------------------------------------------------------------------------------------------------------------%

\begin{figure}
	\centering
		\includegraphics[width=8.cm]{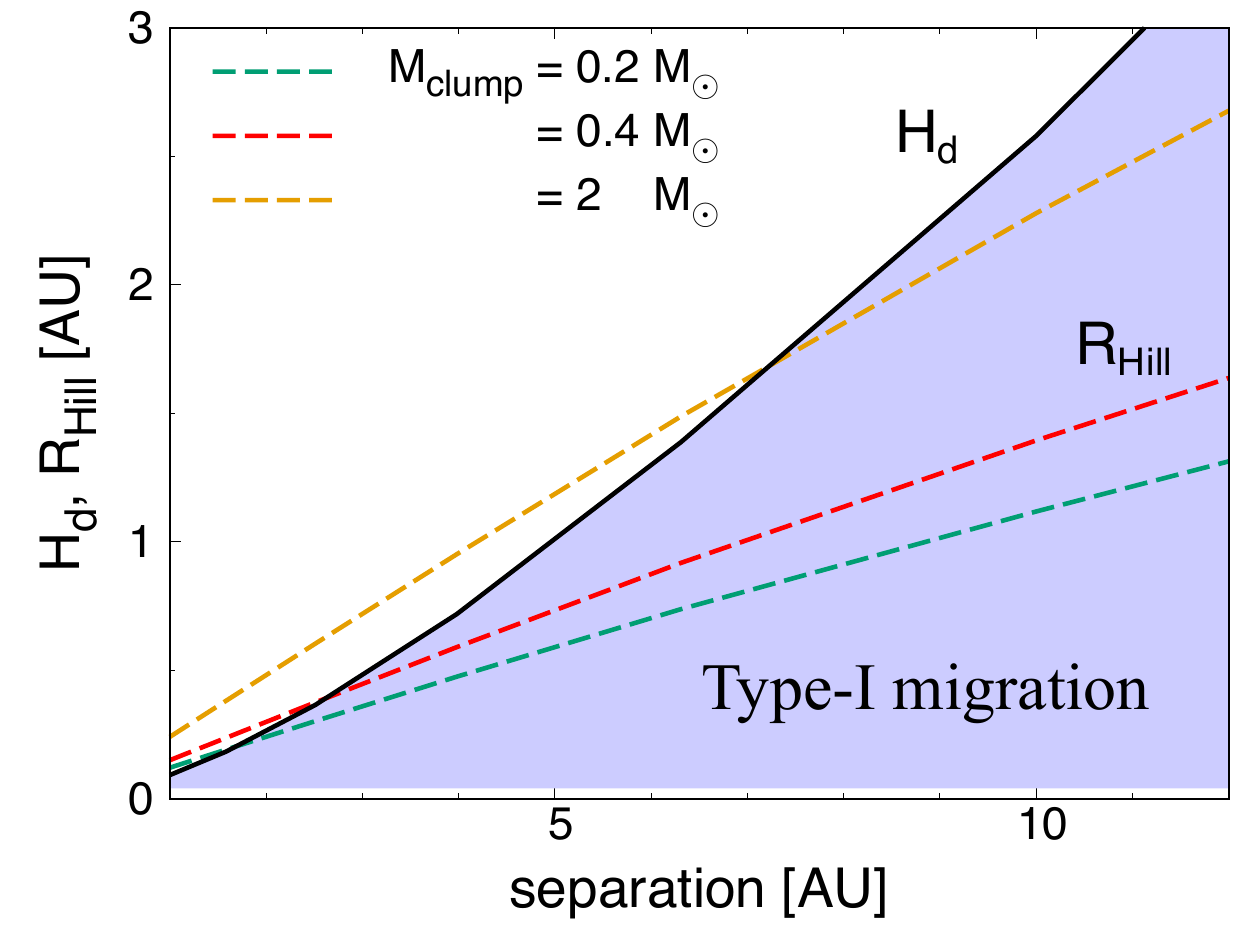}
		\caption{Comparisons between the disk scale height $H_\text{d}$ (solid line) and the Hill radius $R_\text{Hill}$ (dashed line) at different positions apart from the central star. The green, red, and yellow dashed lines represent the Hill radii with the different clump masses of $0.2$, $0.4$, and $2~M_{\odot}$, respectively. To evaluate $H_\text{d}$, we use eq.~\eqref{eq_Hdisk} assuming the mass of the central star is $5.7~M_{\odot}$ and $c_\text{s}=10~\mathrm{km~s^{-1}}$. }
\label{fig_Hd_RHill}
\end{figure}

In Fig.~\ref{fig_Hd_RHill}, we compare the disk scale height $H_\text{d}$ and the Hill radius $R_\text{Hill}$ for the different points apart from the central star.
The green, red, and yellow lines represent $R_\text{Hill}$ with the different clump masses $M_\text{clump} = 0.2$, $0.4$, and $2~M_{\odot}$, respectively.
The figure indicates that the mass accretion onto the clump is important for the gap opening.
Even if we consider the highest examined value of the core mass $0.2~M_\odot$, $R_\text{Hill}$ is always smaller than $H_\text{d}$ except for a very inner part of the disk. However, if the clump has the total mass of $2~M_\odot$ via the accretion, $H_\text{d} < R_\text{Hill}$ is satisfied for $r \lesssim 7$~AU. That is, the gap opening may occur once such a massive clump migrates to enter the inner $r \lesssim 7$~AU region of the disk. The inward migration then stalls afterwards.

%------------------------------------------------------------------------------------------------------------------%

Actually another necessary condition, the so-called ``viscous condition'', is usually considered for the gap opening \citep[e.g.,][]{Crida+2006,DelValle+2012}. It requires that the gap opening timescale $t_\text{open}$ should be smaller than the viscous timescale $t_\text{vis}$, since the gap is filled by the viscous advection flow coming from the gap edges otherwise. Such a condition is written as,
\begin{align}
q > &\frac{243\pi}{8} \alpha \left ( \frac{H_\text{d}}{a} \right )^{2} \nonumber \\
= & 0.95 \left (\frac{\alpha}{0.1} \right ) \left ( \frac{c_\text{s}}{10~\mathrm{km~s^{-1}}} \right )^{2} 
\left ( \frac{a}{10~\mathrm{AU}} \right ) 
\left ( \frac{M_{*}}{10~M_\odot} \right )^{-1},
\end{align}
where $\alpha$ is the viscosity parameter, which typically takes $\alpha = 0.1$--$1$ in self-gravitating disks.
In our simulations, $q$ is always less than $0.1$ even counting the mass accreted during the migration. The above condition is thus not satisfied, meaning that the gas should continuously enter the gap. However, the gap structure is not smeared out in our simulations. 
The gas flowing into the gap is immediately captured by the clump to join its envelope (see Section~\ref{sec_qa_diagram}). We omit the viscous condition to interpret our results below. 

%------------------------------------------------------------------------------------------------------------------------%

\begin{figure}
	\centering
		\includegraphics[width=8.5cm]{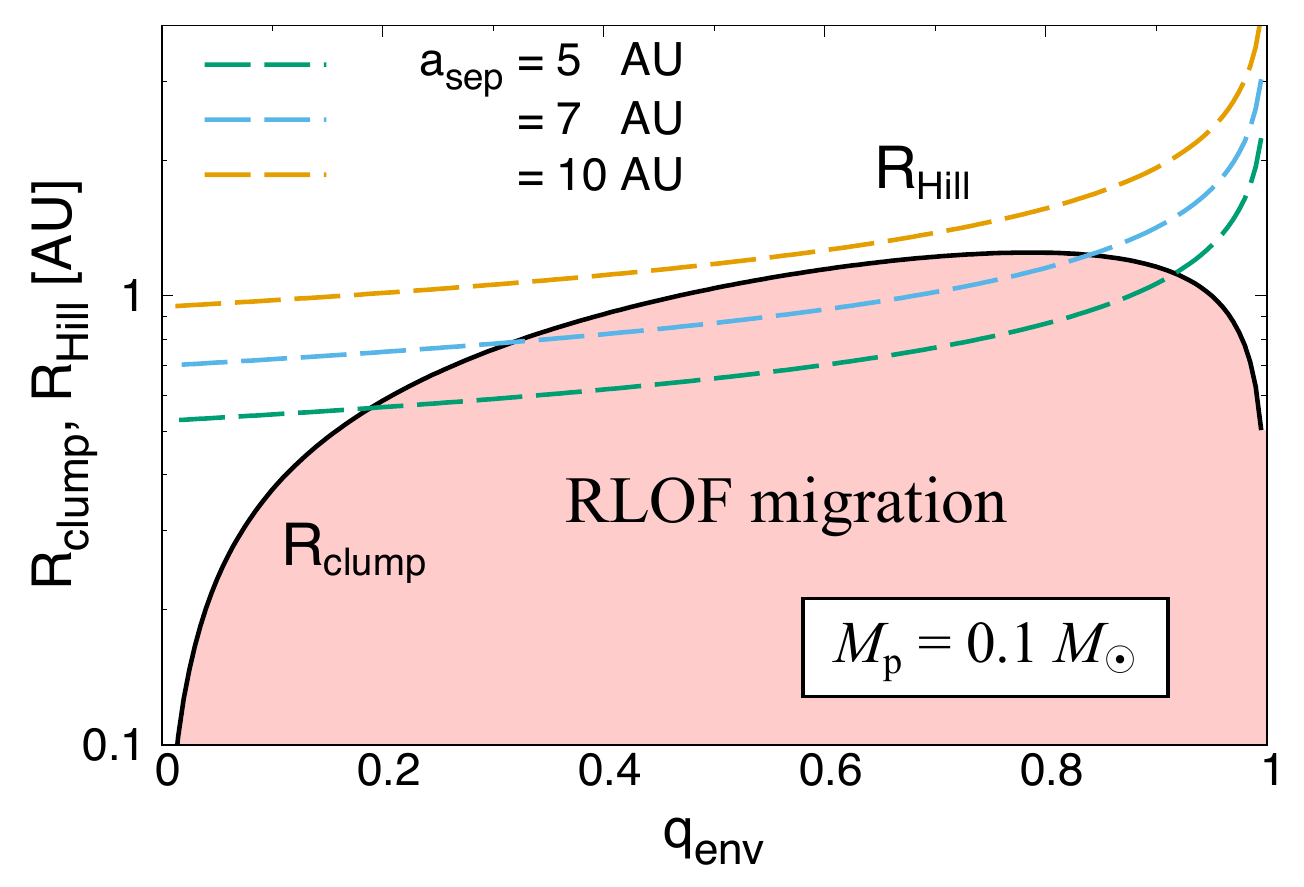}
		\caption{The clump radius $R_\text{clump}$ (solid line) and the Hill radius $R_\text{Hill}$ (dashed lines) as functions of $q_\text{env} \equiv M_\text{env}/M_\text{clump}$, where $M_\text{env}$ is the envelope mass and $M_\text{clump} = M_\text{env} + M_\text{p}$ is the total clump mass. The core mass $M_\text{p}$ is assumed to be $0.1~M_{\odot}$ in this plot. The green, blue, and yellow dashed lines represent the Hill radii with the different binary separations of $5$, $7$, and $10~$AU, respectively.
		}
\label{fig_Rc_RHill}
\end{figure}

\subsubsection{Roche-lobe overflow and further migration} 
\label{sec_Rochelobe}

Our simulations actually suggest that, at least in some cases, the inward migration still continues even in the inner part of the disk, where the above Type-I migration should be inefficient. To explain this, we consider another process which drives the migration: the tidal stripping of the clump envelope \citep{Nayakshin2017rev}. When the clump approaches the central star, the strong tidal force deforms a clump to tear apart an outskirt of the envelope. Some amount of the angular momentum is also lost with the stripped envelope, which may result in the further clump migration. Fortunately, this is similar to the well-studied ``Roche-lobe overflow (RLOF)'' occurring in the binary stellar evolution \citep[e.g.][]{Paczynski1971, Ivanova+2013, Ricker&Taam2012, Passy+2012, Ohlmann+2016, MacLeod+2018}. In what follows, we consider under which conditions such an RLOF-driven migration may occur in our cases. 

%-----------------------------------------------------------------------------------------------------------------------------------------------------------------%

First of all, whether the binary orbit shrinks owing to the RLOF depends on the specific angular momentum of the outflowing gas. \citet{Huang1963a} shows that the binary separation shrinks if the following condition is satisfied,
\begin{align}
\gamma_\text{loss} \equiv \frac{h_\text{loss}}{h_\text{bin}} > \frac{1}{2} + \frac{M_*}{M_\text{clump}} \sim \frac{M_*}{M_\text{clump}},
\end{align}
where $h_\text{loss}$ and $h_\text{bin}$ are the specific angular momentum of the overflowing gas and that of the binary, $M_\text{clump}$ and $M_*$ are the masses of the migrating clump and the central star. We apply the above condition to our cases neglecting the first term, because $M_*$ is larger by an order of magnitude than $M_\text{clump}$.
Since the outflowing gas has $\gamma_\text{loss} \gtrsim 2 M_{*}/M_\text{clump}$ in our calculation, 
which is larger than the above critical value, 
the binary separation decreases when RLOF occurs.

%-------------------------------------------------------------------------------------------------------------------------------------------------------------%

Once the clump size $R_\text{clump}$ exceeds the Hill radius $R_\text{Hill}$, an outer part of the envelope becomes no longer gravitationally bound. In order to consider when the RLOF starts satisfying $R_\text{clump} > R_\text{Hill}$, we analytically model the clump structure. Since the hydrostatic balance is achieved in the clump envelope, the structure is approximately determined by the Lane-Emden equation. The so-called ``M-solutions'' found by \citet{Chandrasekhar1939} represent the structure of a gas envelope bound by a condensed central core \citep{Osterbrock1953,Hjellming+1987}. Such M-solutions with the adiabatic exponent $\gamma = 5/3$ are reasonable models of the migrating clumps, providing an analytic expression of the envelope radius $R_\text{clump}$,
\begin{align} 
\label{eq_rclump}
R_\text{clump} &= 1.2~\mathrm{AU} \nonumber \\
&A(q_\text{env}) \left ( \frac{T_{0}}{8000~\mathrm{K}} \right ) \left ( \frac{n_\text{adib}}{10^{16}~\mathrm{cm^{-3}}}\right )^{-2/3}
\left ( \frac{M_\text{p}}{0.1~M_{\odot}}\right )^{-1/3},
\end{align}
where $q_\text{env}$ is the mass fraction of the envelope to the migrating clump, defined as $M_\text{env} / M_\text{clump}$, where $M_\text{env}$ is the mass of the envelope. The non-dimensional function $A(q_\text{env})$ assures that an envelope solution is connected to the central core satisfying appropriate boundary conditions. 

%-----------------------------------------------------------------------------------------------------------------------------------------%

The solid line in Fig.~\ref{fig_Rc_RHill} represents the clump radius $R_\text{clump}$ given by the M-solutions. We see that $R_\text{clump}$ reaches its maximum around $q_\text{env} \simeq 0.7$. The radius $R_\text{clump}$ decreases with decreasing $q_\text{env}$ for $q_\text{env} \lesssim 0.7$.
In this regime, the gravity of the central core becomes dominant over the self-gravity of the envelope. The envelope is tightly bound by the core to have the small radius in the lower end of $q_\text{env}$. In the higher end of $q_\text{env}$, on the other hand, $R_\text{clump}$ decreases with increasing $q_\text{env}$. The self-gravity of the envelope becomes much stronger than the gravity from the core in this case. 

%--------------------------------------------------------------------------------------------------------------------------------------------%

Fig.~\ref{fig_Rc_RHill} shows that $R_\text{clump} > R_\text{Hill}$ can be satisfied only for the smaller binary separation $a$. For $a=10~$AU, for instance, $R_\text{clump}$ is always smaller than $R_\text{Hill}$ for any values of $q_\text{env}$. With the smaller separation $a= 7~$AU, however, $R_\text{clump} > R_\text{Hill}$ is satisfied for $0.3 \lesssim q_\text{env} \lesssim 0.8$, meaning that the RLOF occurs. As the separation $a$ decreases owing to the accompanying inward migration, $R_\text{clump} > R_\text{Hill}$ is more easily satisfied except for the limiting cases with $q_\text{env}$ very close to $0$ or $1$. That is, the RLOF-driven migration should be accelerated once the clump enters an inner part of the disk, taking a moderate value of $q_\text{env}$.

%-----------------------------------------------------------------------------------------------------------------------------------------------%
\begin{figure*}
	\centering
		\includegraphics[width=16.1cm]{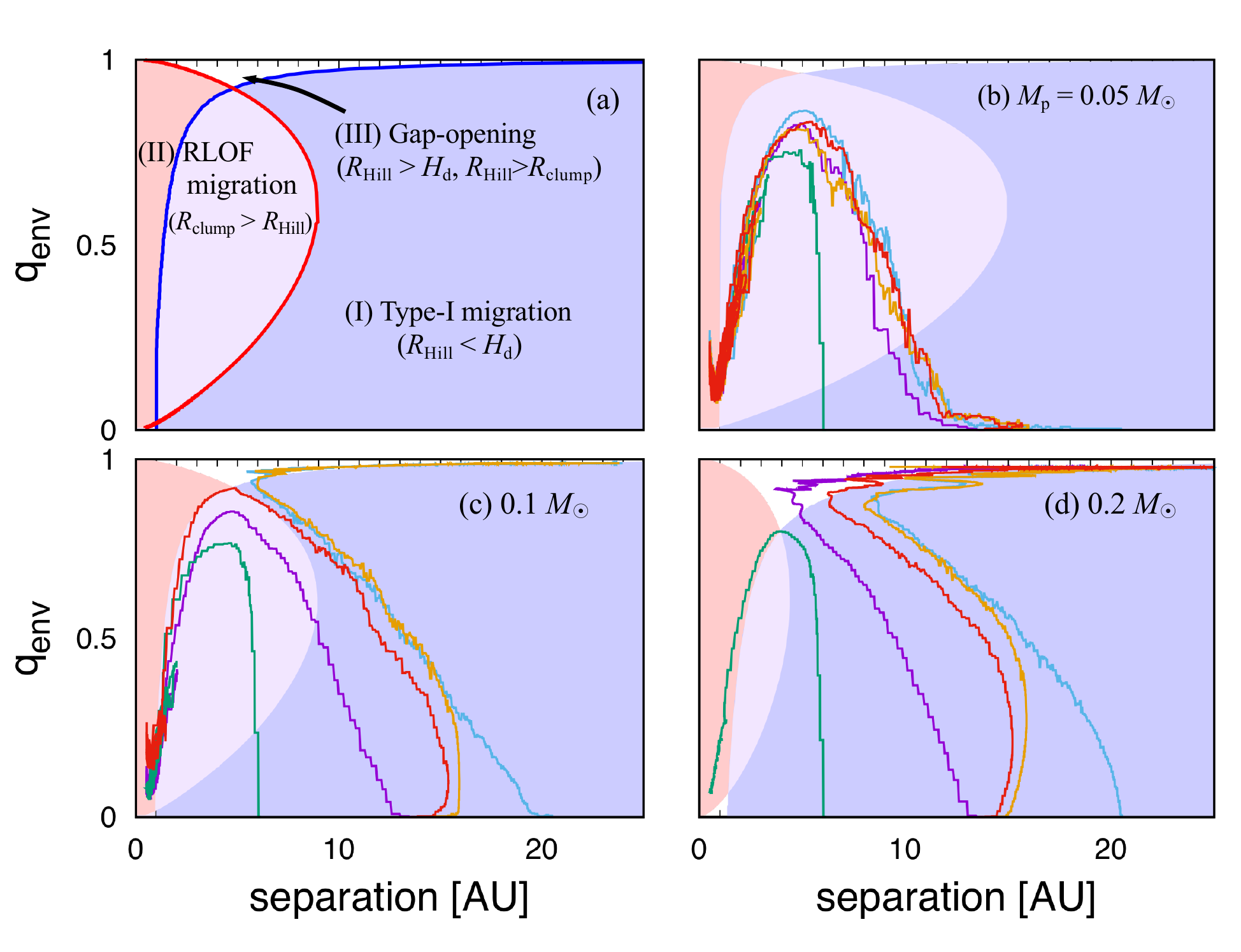}
		\caption{(a) A summary of the analytic considerations on the clump migration through disks. The horizontal and vertical axes represent the mass ratio $q_\text{env} = M_\text{env}/M_\text{clump}$ and the separation between the central star and clump.
The blue and red lines are boundaries where the critical conditions $R_\text{Hill} = H_\text{d}$ and $R_\text{clump} = R_\text{Hill}$ are satisfied, respectively. The inward migration is expected in the color-shaded areas; the Type-I migration below the blue line where $R_\text{Hill} < H_\text{d}$, and RLOF-driven migration in the left-hand-side of the red line where $R_\text{clump} > R_\text{Hill}$. The migration is avoidable only above both the blue and red lines, where $R_\text{Hill} > H_\text{d}$ and $R_\text{clump} < R_\text{Hill}$ are satisfied. (b) Comparisons between the analytic considerations and numerical results for $M_\text{p} = 0.05~M_\odot$. The different evolutionary tracks represent the cases with the different initial positions of A (purple), B (green), C (yellow), D (blue), and E (red). 
(c,d) Same as panel (b) but for the different initial core masses of $M_\text{p} = 0.1$ and $0.2~M_\odot$.
}
\label{fig_rq_evolution}
\end{figure*}
%--------------------------------------------------------------------------------------------------------------------------%

%%%%%%%%%%%%%%%%%%%%%%%%%%%%%%%%%%%%%%%%%%%%%%%%%%%%%%%%%%%%%%%%%%%%%%%%%
\subsection{Simulation Results {\it versus} Analytic Considerations} 
\label{sec_qa_diagram}
%%%%%%%%%%%%%%%%%%%%%%%%%%%%%%%%%%%%%%%%%%%%%%%%%%%%%%%%%%%%%%%%%%%%%%%%%

In this section, we compare our simulation results to the analytic considerations, which suggest that the gas accretion onto a migrating clump should play a key role to determine its fate. We quantify the gas mass accreted onto the clump core using the variable $q_\text{env} \equiv M_\text{env} / M_\text{clump}$ below. 

%-----------------------------------------------------------------------------%

Fig.~\ref{fig_rq_evolution}~(a) summarizes the analytic considerations in the parameter space of $q_\text{env}$ and $a$. The color-shaded areas represent that the efficient inward migration is expected because the orbital angular momentum can be extracted.
For instance, the condition $H_\text{d} < R_\text{Hill}$ is satisfied in the area below the blue curve, indicating that the inward Type-I migration should occur (see Section~\ref{sec_gapformation}). The other condition $R_\text{clump} < R_\text{Hill}$ for the migration is satisfied in the area enclosed by the red curve, where the RLOF drives the migration (see Section~\ref{sec_Rochelobe}).

%------------------------------------------------------------------------------%

\begin{figure}
	\centering
		\includegraphics[width=8.1cm]{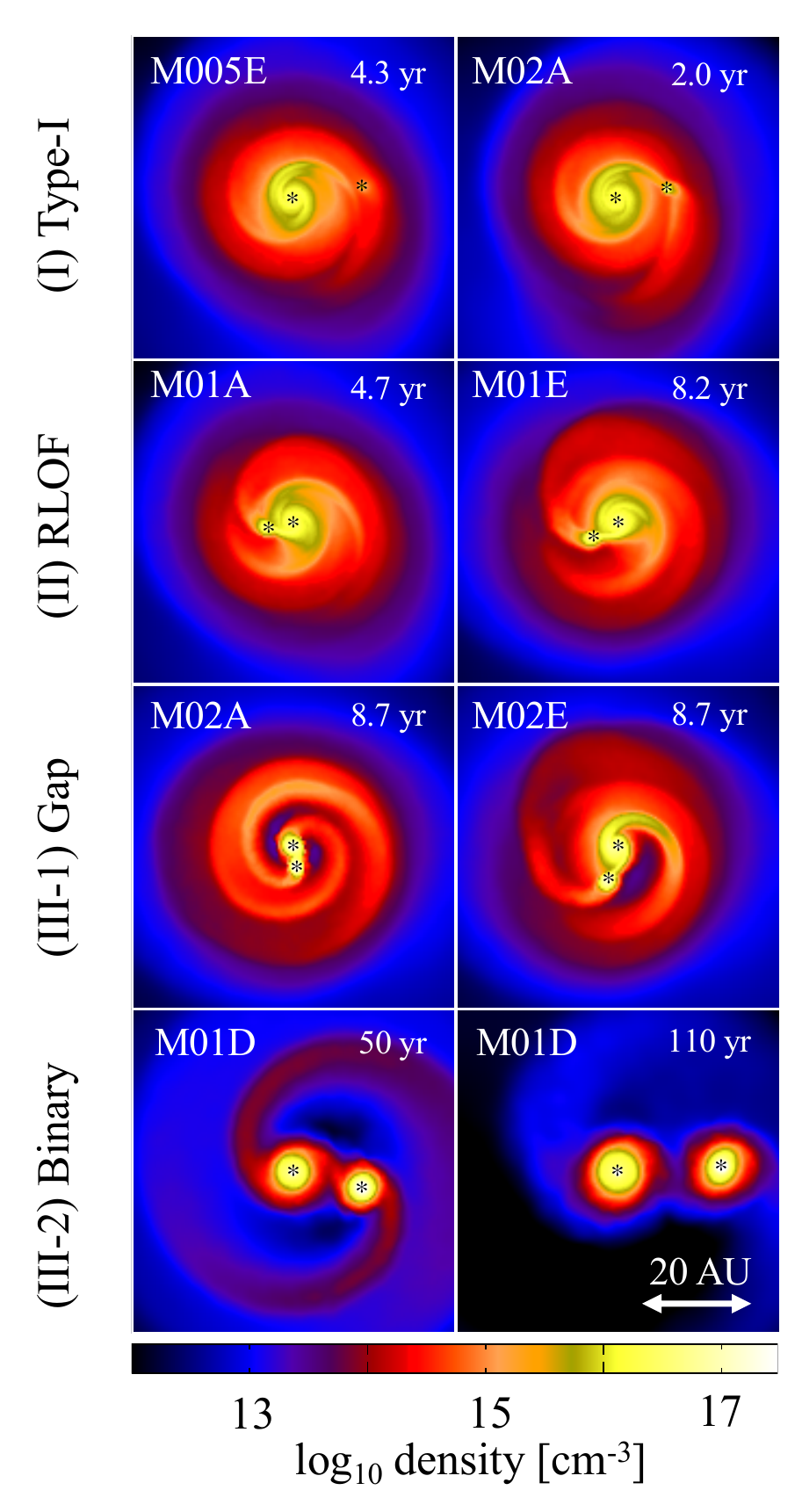}
		\caption{The typical density distributions in the disk midplane for the different evolutionary stages shown in Fig.~\ref{fig_rq_evolution}. The two representative snapshots are presented for each evolutionary stage of (I) Type-I migration, (II) RLOF-driven migration, (III-1) Gap-opening, and (III-2) Binary, from the top to bottom rows. The model name and elapsed time after inserting the point particle are presented in each panel. }
		\label{fig_map_phases}
\end{figure}

%------------------------------------------------------------------------------%

Fig.~\ref{fig_rq_evolution}~(b), (c), and (d) show the trajectories of the clumps in our simulations with the different initial masses of $0.05$, $0.1$, and $0.2~M_\odot$. In each panel, the different lines represent the different initial positions. The background colors denote the analytic predictions in the same way as in panel (a). Overall, the divergent evolution in our numerical simulations are well explained by the analytic considerations. In all the cases, the separation initially decreases owing to the Type-I migration. The mass ratio $q_\text{env}$ increases during that, representing the clump mass growth via the accretion.
Fig.~\ref{fig_rq_evolution}~(b) shows that the migration continues until the clumps merge with the central star for all the cases. We see that $q_\text{env}$ turns to decrease for $a < 5~\text{AU}$, indicating that the clumps loose their masses via the tidal disruption of their envelopes.

%--------------------------------------------------------------------------------%

Fig.~\ref{fig_rq_evolution}~(c) and (d) also show that $q_\text{env}$ initially rises with decreasing the separation. However, some trajectories show the qualitatively different subsequent evolution; the inward migration stops immediately after passing through the blue curve, above which $R_\text{Hill}$ is larger than $H_\text{d}$. 
The gap opening occurs at this moment (stage III-1 in Fig.~\ref{fig_map_phases}), as predicted by the analytic consideration. Moreover, the separation turns to increase after that. Fig.~\ref{fig_rq_evolution} shows that the clump trajectories evolve along the boundary of $H_\text{d} = R_\text{Hill}$. The clump and central star are in a binary system during that (stage III-2 in Fig.~\ref{fig_map_phases}).

%--------------------------------------------------------------------------------%

\begin{figure}
	\centering
		\includegraphics[width=8.5cm]{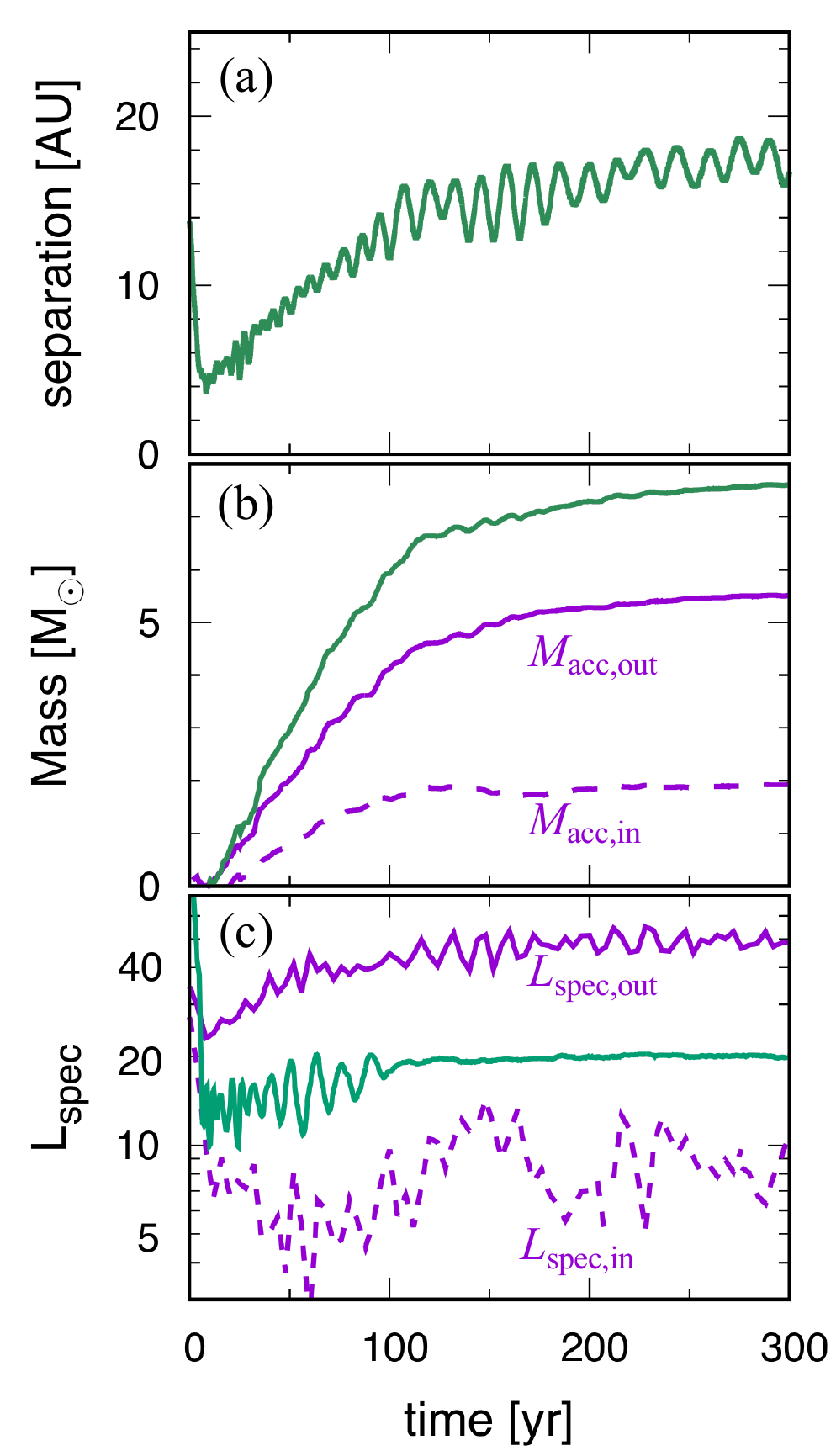}
		\caption{
		Mass and angular momentum accretion histories during the later outward migration for case M02A.
		(a) Time evolution of the binary separation.
		(b) The mass growth histories of the clump. The purple solid and dashed lines represent the gas mass accreted from the outer and inner part of the disk (inside and outside of the binary orbit), and the green line represents the sum of them.
		We normalize the mass to be 0 at $t=10~\mathrm{years}$, when the binary separation turns to increase.
		(c) The specific angular momentum of the accreting gas. The purple solid and dashed lines represent those of the gas coming from the outer and inner part of the disk. The green line shows the time variation of the specific angular momentum of the clump.
		}
		\label{fig_Mass_Inflow}
\end{figure}

In order to investigate what causes the outward migration, we analyze how the clump acquires the angular momentum. Fig.~\ref{fig_Mass_Inflow} shows such an example for case M02A (purple line in Fig.~\ref{fig_rq_evolution}~d). The clump and central star evolve being in a binary system for $t \gtrsim 10~$years, during which the outward migration occurs. 
As shown in Fig.~\ref{fig_Mass_Inflow}~(a), the migration continues until it gradually ceases for $t \gtrsim 100~$years. 
Fig.~\ref{fig_Mass_Inflow}~(b) and (c) indicate that the mass accretion onto the clump is a key to understand the evolution. As shown in panel (b), in fact, the mass growth via the accretion ceases for $t \gtrsim 100~$years, which is synchronous to the migration rate. We further analyze how the accreting gas brings the orbital angular momentum into the clump. To this end, we divide the accreted gas into two components; the gas coming from the inside and outside of the binary orbit.
The purple lines in panel (b) represent the masses brought by these two components.
We see that the larger amount of the gas comes from the outer part of the disk.
This component has larger specific angular momentum than the migrating clump (Fig~\ref{fig_Mass_Inflow}~c).
Such an ``outside-dominant'' accretion widens the binary orbit.
Similar phenomena have been reported in recent studies on both the planetary and BH migrations \citep[e.g.,][]{Stamatellos+2018, Munoz+2019}.

%%%%%%%%%%%%%%%%%%%%%%%%%%%%%%%%%%%%%%%%%%%%%%%%%%%%
\subsection{Limits on the outward migration} 
\label{sec_final_separation}
%%%%%%%%%%%%%%%%%%%%%%%%%%%%%%%%%%%%%%%%%%%%%%%%%%%%

\begin{figure}
	\centering
		\includegraphics[width=8.7cm]{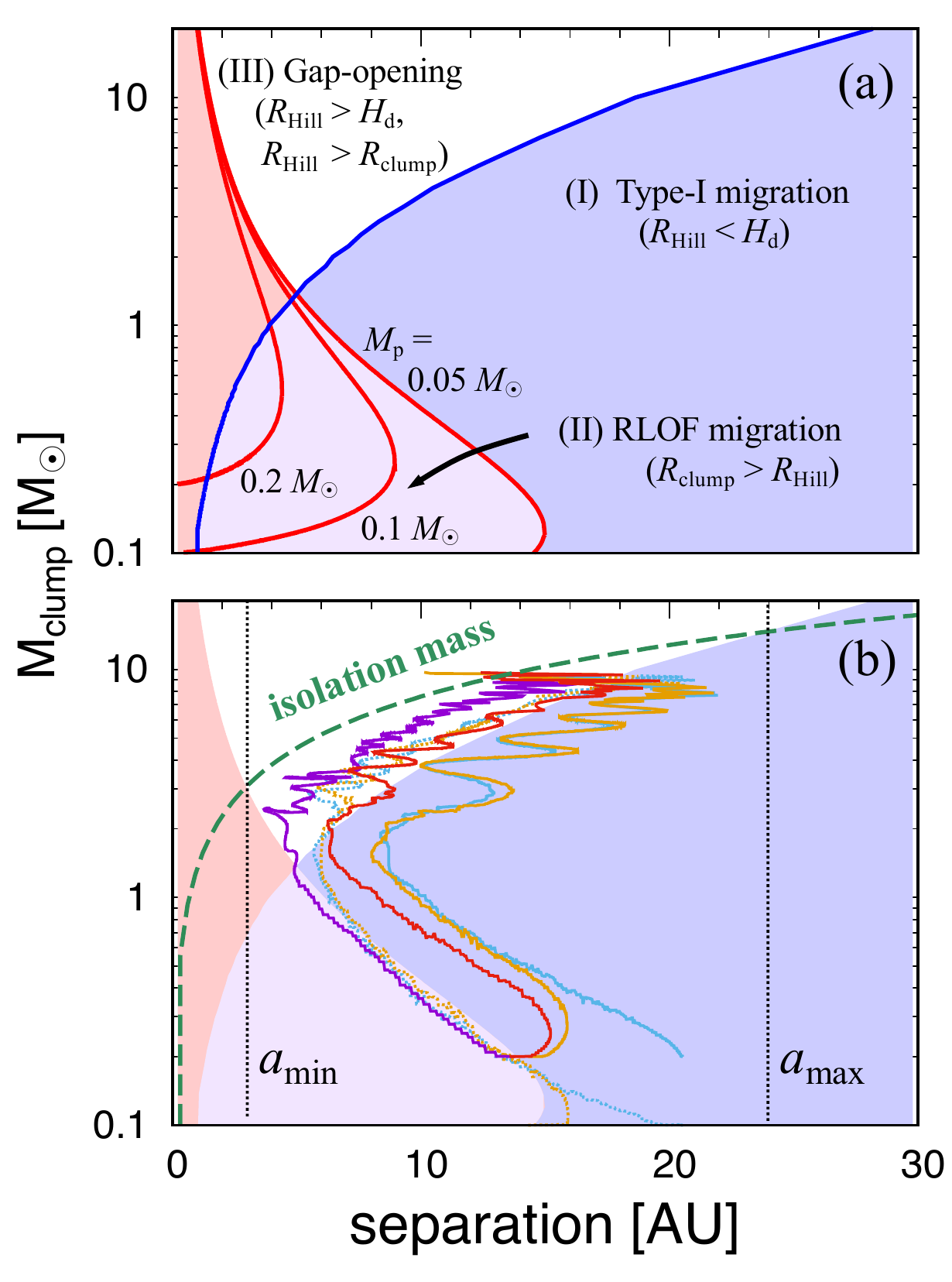}
		\caption{
Similar to Fig.~\ref{fig_rq_evolution} except for using the clump mass for the vertical axes. (a) A summary of the analytic considerations on the clump migration through disks.		
		Inside the shaded regions, the clump can migrate inward by the spiral arms excited on the disk
		(Type-I migration; blue) and the tidal force from the central star (RLOF migration; red).
		The solid lines represent the boundaries where the above migration process operates.
		The different red lines represent the different core masses assumed, $M_\text{p} = 0.05$, $0.1$, and $0.2~M_\odot$.
%%%%%%%%
		(b) Comparisons between the analytic considerations and numerical results. The dotted and solid trajectories represent the numerical results with the different initial core masses of $0.1$ and $0.2~M_\odot$. The different colors represent the different initial positions A (purple), C (yellow), D (blue), and E (red). The green dashed line represents the isolation mass given by eq.~\eqref{eq_Miso} assuming the marginally stable self-gravitating disks with $Q = 1$. The vertical black dotted lines indicate the maximum and minimum separations given by eqs. \eqref{eq_a_max} and \eqref{eq_a_min}.
		}
		\label{fig_rm_plane}
\end{figure}

The above analyses suggest that the outward migration should continue as far as a clump accretes the disk gas. However, it also means that the migration stalls when the clump cannot accrete a sufficient amount of the gas to drive it. We investigate such possible limits below. 

%---------------------------------------------------------------------------------%

The mass accretion is terminated when a clump has accreted most of the disk gas near the orbit. Such a limiting mass is analytically estimated as follows \citep[e.g.,][]{Zhu+2012, Inayoshi+2014b,Latif+2015b,Latif+2015c}. 
Consider the clump in a circular orbit with the radius $R$ in a self-gravitating disk. The disk is assumed to be marginally gravitationally unstable with $Q = 1$, which is the case in our simulations.
Since the clump only attracts the gas within the Hill radius, maximum mass of the gas possibly bound by the gravity of the clump is written as
\begin{align} 
M_\text{iso} (R) &= 4 \pi R\; R_\text{Hill} \Sigma (R) \nonumber \\
 &= M_{*} \left ( \frac{64}{3Q^{3}} \right )^{1/2} \left ( \frac{c_\text{s}}{v_\text{Kep}} \right ) ^{3/2} \nonumber \\
 &= 9~M_{\odot} \left ( \frac{M_{*}}{10~M_{\odot}} \right ) ^{1/4} \left ( \frac{T}{10^{4}~\mathrm{K}} \right )^{3/4}
 \left ( \frac{R}{10~\mathrm{AU}}\right )^{3/4},\label{eq_Miso}
\end{align}
where $M_\text{iso}$ stands for the ``isolation mass'' \citep[e.g.,][]{Lissauer1987,Goodman&Tan2004}. Since the Hill radius increases to cover the larger part of the disk for the larger $R$, $M_\text{iso} (R)$ becomes comparable to the disk mass enclosed within $R$.

%-------------------------------------------------------------------------------------------------------------%

Fig.~\ref{fig_rm_plane} is similar to Fig.~\ref{fig_rq_evolution} except for using the clump mass $M_\text{clump}$ for the vertical axes. As in Fig.~\ref{fig_rq_evolution}~(a), panel (a) summarizes the analytic considerations on the evolution of the migrating clumps. The red and blue lines represent the critical boundaries of $R_\text{clump} = R_\text{Hill}$ and $R_\text{Hill} = H_\text{d}$, respectively. Note that the different red lines correspond to the different core masses $M_\text{p} = 0.05, 0.1,$ and $0.2~M_{\odot}$. The inward migration is avoidable only in the area above both the red and blue lines, where a gap is cleared around the clump within the disk. 

%-------------------------------------------------------------------------------------------------------------%

In Fig.~\ref{fig_rm_plane}~(b), we plot the clump trajectories for the cases where the later outward migration occurs. We see that the trajectories go along the boundary of $R_\text{Hill} = H_\text{d}$ during the outward migration. 
In panel (b), the dashed line represents the isolation mass $M_\text{iso}$ given by eq.~ \eqref{eq_Miso}. The trajectories do not exceed the line of $M_\text{iso}$, by which the clump mass growth is limited. Therefore, there is the maximum separation $a_\text{max}$, beyond which the outward migration is not possible without the further mass growth. 
Using eqs.~\eqref{eq_Hdisk}, \eqref{eq_rHill}, and \eqref{eq_Miso}, we obtain the scaling relation
\begin{equation}
 a_\text{max} \simeq 20~\mathrm{AU} \left( \frac{M_{*}}{5.7~M_{\odot}} \right ) \left ( \frac{c_\text{s}}{10~\mathrm{km~s^{-1}}} \right )^{-2} , 
\label{eq_a_max}
\end{equation}
where $M_{*}$ is the mass of the central star.
In our simulations, in fact, the clumps cease to migrate outward before the separation reaches $20$~AU.

%---------------------------------------------------------------------------------------------------------------%

Fig.~\ref{fig_rm_plane}~(b) also suggests that the early inward migration is never reversed after crossing the minimum separation $a_{\rm min}$.
Recall that the RLOF-driven migration operates near the central star, unless a clump has a sufficiently massive envelope. 
The critical condition for that is represented by the red line, below which the RLOF-driven migration is possible to occur. 
However, the available gas mass is again limited by the isolation mass represented by the green dashed line. The minimum separation is thus given by the crossing point between these lines. Using eqs.~\eqref{eq_Hdisk}, \eqref{eq_rclump}, and \eqref{eq_Miso}, we get
\begin{align}
a_\text{min}  &\simeq 3~\mathrm{AU}   \nonumber \\
&\;\;  \left( \frac{M_{*}}{5.7~M_{\odot}} \right )^{1/9} 
\left ( \frac{c_\text{s}}{10~\mathrm{km~s^{-1}}} \right )^{2/3} \left ( \frac{n_{0}}{10^{16}~\mathrm{cm^{-3}}} \right )^{-4/9} .
\label{eq_a_min}
\end{align}
Note that the value of $a_\text{min}$ is almost independent of the initial core mass $M_\text{p}$. This is because, for a massive clump, the self-gravity of the envelope mainly contributes to determine the radius rather than that of the core.
Another interesting fact is that $a_\text{min}$ is close to the Jeans length for a wide mass range. Eq.~\eqref{eq_a_min} is actually transformed as
 \begin{align}
a_\text{min}  &\simeq 2~\mathrm{AU}   \nonumber \\
&\;\;  \left( \frac{M_{*}}{M_\text{Jeans}} \right )^{1/9} 
\left ( \frac{c_\text{s}}{10~\mathrm{km~s^{-1}}} \right ) \left ( \frac{n_{0}}{10^{16}~\mathrm{cm^{-3}}} \right )^{-1/2} ,\nonumber \\
& \simeq \left( \frac{M_{*}}{M_\text{Jeans}} \right )^{1/9} R_\text{Jeans},
\label{eq_a_min_jeans}
\end{align}
where $R_\text{Jeans}$ and $M_\text{Jeans}$ are the Jeans length and mass.
We can see that $a_\text{min}$ is close to the Jeans scale
due to the weak dependence on its mass.
%------------------------------------------------------------------------------------%

\begin{figure}
	\centering
		\includegraphics[width=8.cm]{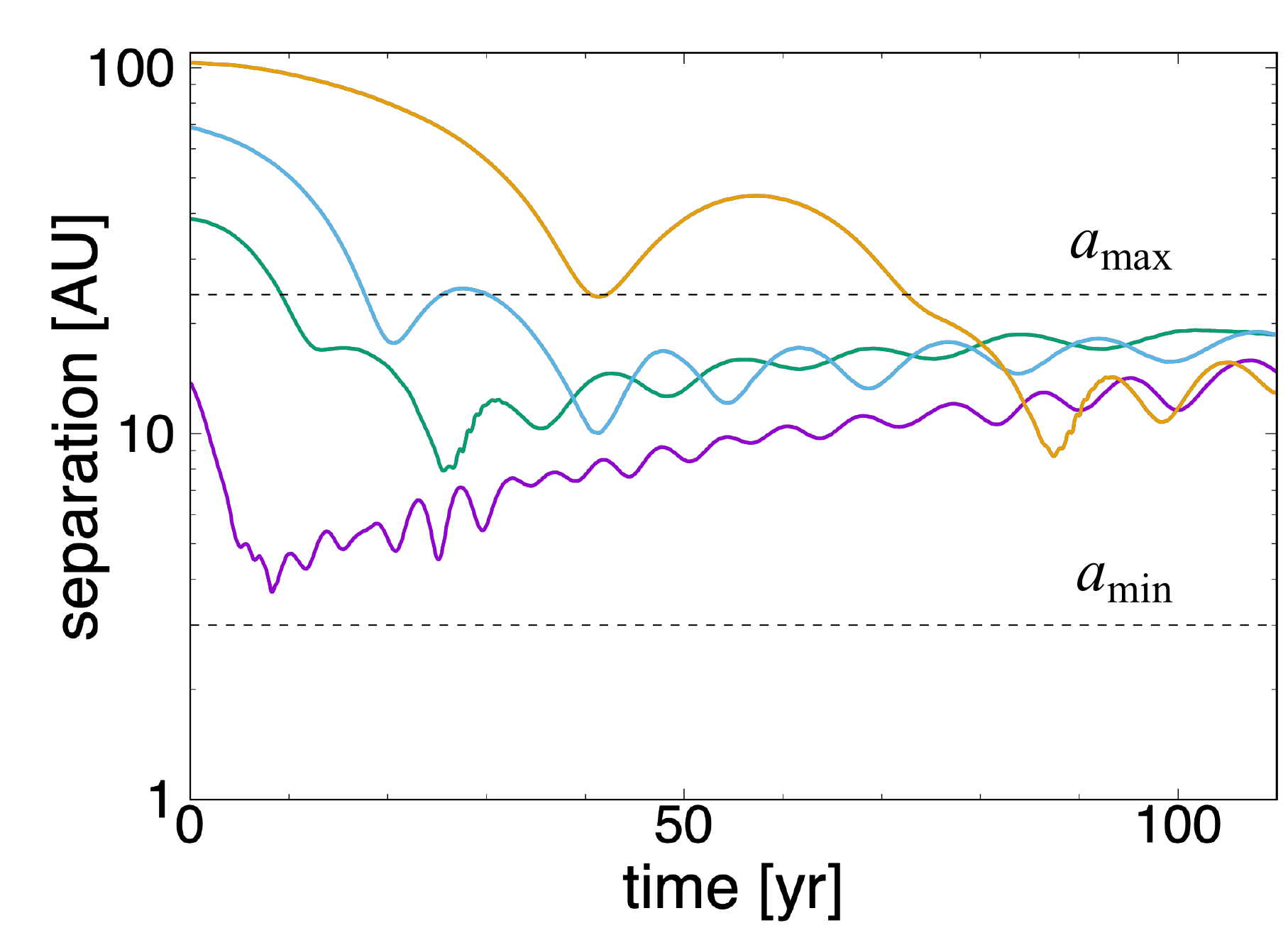}
		\caption{Time evolution of the separations between the central star and clumps with the large initial separations. Presented are only the cases with the initial core mass $0.2~M_{\odot}$. Regardless of the variation of the initial separations, the separations finally converge to $10$--$20~$AU for all the examined cases. The horizontal dotted lines represent the maximum and minimum separations given by eqs. \eqref{eq_a_max} and \eqref{eq_a_min}.}
		\label{fig_r_t_evolution_max}
\end{figure}

%-----------------------------------------------------------------------------------%

To reinforce the above arguments, we further perform additional numerical experiments. We follow the orbital evolution with the large initial separations from the central star, $15, 40, 70,$ and $105~$~AU for the fixed initial core mass $0.2~M_\odot$. The typical disk size during the evolution is $\simeq 30~$AU, and the initial clump positions are far outside of the disk except for the case with $15$~AU. Fig.~\ref{fig_r_t_evolution_max} shows the orbital evolution for these models. In all the models, the early inward migration is reversed to the outward migration, after which the separations finally converge to $\simeq 20~$AU. This value is consistent with the maximum separation given by eq.~\eqref{eq_a_max}. 
We note that the final separation is not just determined by the disk size, though they are comparable by chance for the cases examined here. In Appendix~\ref{sec_revo_sd}, we further investigate similar cases where the disk mass is much smaller than the above. The outward migration ceases near the maximum radius $a_\text{max}$, which is far smaller than the disk size.

%%%%%%%%%%%%%%%%%%%%%%%
%%%%%%%%%%%%%%%%%%%%%%%
\section{Discussion} 
\label{sec_discussion}
%%%%%%%%%%%%%%%%%%%%%%%
%%%%%%%%%%%%%%%%%%%%%%%

\subsection{Equation of State and Cooling}

Our analyses have shown that the clump radius $R_\text{clump}$ is an important quantity to determine the fate of its orbital evolution. 
The analytic description of $R_\text{clump}$ (eq.~\ref{eq_rclump}) is derived assuming 
 the adiabatic EOS for the envelope gas. 
This reproduces the clump radii found in our numerical experiments,
where the adiabatic EOS is also assumed at the density $n > n_\text{adib}$.
Our assumption of the EOS supposes the fact that the gas is optically thick to the local radiation field at such high densities.
%-----------------------------------------------------------------------------------%

In reality, however, the gas should not be perfectly adiabatic. A clump in the hydrostatic balance looses energy by radiating away from the surface.  
Such a clump undergoes the quasi-static contraction, often referred to as the Kelvin-Helmholz (KH) contraction.
The timescale for the contraction is given by the so-called KH timescale
\begin{align}
t_\text{KH} = \frac{GM_\text{clump}^{2}}{R_\text{clump} L_\text{cool}},
\end{align}
where $M_\text{clump}$ is the clump mass and $L_\text{cool}$ is the surface luminosity. The clump radius can deviate from that given by eq.~\eqref{eq_rclump} over the above timescale. Our assumption of the adiabatic EOS is reasonable if the KH time is much longer than the clump orbital time. Otherwise, the clump size would be reduced owing to the radiative cooling during the early inward migration. The clump would have a large chance of the survival, because the tidal disruption and resulting RLOF migration become difficult to occur for a compact object. 

%------------------------------------------------------------------------------------%

To confirm the validity of the adiabatic EOS for our DC cases, we here compare the KH and orbital timescales. The clump surface luminosity $L_\text{cool}$, which is necessary to estimate the KH timescale, depends on the detailed interior structure.
Using the results of numerical modelling of the accreting primordial stars by \citet{Hosokawa+2012},
\begin{align}
L_\text{cool} &\simeq 0.6~L_{\odot} \left ( \frac{M_\text{clump}}{M_{\odot}} \right )^{11/2}  
\left ( \frac{R_\text{clump}}{R_{\odot}} \right )^{-1/2},
\end{align}
these timescales are estimated as
\begin{align}
t_\text{KH} &\simeq 3\times 10^{6} ~\mathrm{yr}  \left ( \frac{M_\text{clump}}{M_{\odot}} \right )^{-7/2}
\left ( \frac{R_\text{clump}}{\mathrm{AU}} \right )^{-1/2}, \label{eq_tcool_prim} \\
t_\text{orb} &= 1.6~\mathrm{yr} \left ( \frac{M_*}{10~M_\odot} \right )^{-1/2} \left ( \frac{a}{10~\mathrm{AU}} \right )^{3/2}.
\end{align}
Since the KH timescale is much longer than the orbital time, the effect of the radiative cooling is totally negligible in our cases. 

%------------------------------------------------------------------------------------%

\subsection{Applications to other star formation channels} 
\label{sec_scale}

As mentioned in Section~\ref{sec_methodology}, our results can be applied to other star formation channels beyond the DC model. Specifically, the governing equations are invariant under the following transformations,
\begin{align} \label{eq_scaling}
\left \{
\begin{array}{llllllll}
c_\text{s} & \longrightarrow &\zeta c_\text{s} &, & \rho &\longrightarrow& \eta \rho &, \\
t & \longrightarrow & \eta^{-1/2} t &, & r & \longrightarrow & \eta^{-1/2} \zeta r &,\\
m & \longrightarrow & \eta^{-1/2} \zeta^{3} m &,&n & \longrightarrow & \eta n&, 
\end{array}
\right.
\end{align}
where $\eta$ and $\zeta$ are non-dimensional scaling parameters and $\rho$, $t$, $r$, $m$, and $n$ are the density, time, distance, mass, and number density.
Note that the maximum and minimum separations given by eqs.~\eqref{eq_a_max} and \eqref{eq_a_min} are invariant under the above manipulation. Below we mostly discuss implications of these equations.

%-----------------------------------------------------------------------------------------------%
\subsubsection{Normal Pop~III star formation}

As described in Section~\ref{sec_intro}, the disk fragmentation is broadly expected in the normal Pop III star formation. Owing to the H$_2$ molecular cooling, a cloud initially collapses with the adiabatic exponent $\gamma = 1.1$ \citep[e.g.,][]{OmukaiNishi1998}.
At the density $n \gtrsim 10^{16} \mathrm{cm^{-3}}$, the gas becomes optically thick against the cooling radiation and $\gamma$ approaches $5/3$ as the density increases
\citep[e.g.][]{Greif+2012}.
Applying eqs.~\eqref{eq_a_max} and \eqref{eq_a_min} to such a case, the minimum and maximum separations are written as,
\begin{align}
a_\text{min}  &= 1.3~\mathrm{AU}   \nonumber \\
&\;\;  \left( \frac{M_{*}}{10~M_{\odot}} \right )^{1/9} 
\left ( \frac{c_\text{s}}{5~\mathrm{km~s^{-1}}} \right )^{2/3} \left ( \frac{n_{0}}{10^{16}~\mathrm{cm^{-3}}} \right )^{-4/9} ,
\label{eq_minimum_sep_popIII} \\
a_\text{max} &= 140~\mathrm{AU}    \left( \frac{M_{*}}{10~M_{\odot}} \right ) \left ( \frac{c_\text{s}}{5~\mathrm{km~s^{-1}}} \right )^{-2}.
\label{eq_maximum_sep_popIII}
\end{align}
These suggest that the separation of the Pop~III binaries ranges from $\sim$AU to $\sim 100~$AU.
We note that the minimum separation has such a weak dependence of $a_\text{min} \propto M_*^{1/9}$, while the maximum separation is in proportion to $M_*$.

%----------------------------------------------------------------------------------%

\citet{Greif+2012} investigate the evolution in the earliest phase of the protostellar accretion stage, performing 3D simulations with a very high spatial resolution down to $\sim 10 ~R_{\odot}$. They observe the disk fragmentation and follow their orbital evolution for a few orbital time. Some clumps formed via the disk fragmentation survive for $\sim 10$~years, which is the duration they follow, with separations of $\sim$~AU from the most massive star. Since the mass of the primary star is $M_* \sim 0.1$--$1~M_\odot$ during the evolution, the separations coincide with our above estimates. Note that in their calculation, the primary star has a radius of $\sim$ AU, which is close to $a_\text{min}$. In this case, the minimum separation is also limited by the stellar radius, below which the clump just merges away. 
Since the size of a newly-born star is close to the Jeans length, $a_\text{min}$ coincides with the stellar radius in their cases (see eq.~\ref{eq_a_min_jeans}).

%----------------------------------------------------------------------------------------------%

Binary properties at larger scales are also studied by several authors \citep[e.g.,][]{Smith+2011,Clark+2011,Susa+2014,Stacy+2016,Hosokawa+2016}, who have followed the longer-term evolution with the lower spatial resolutions. 
They find binaries with typical separations of $100$--$300~$AU, which are comparable to the maximum separation given by eq.~\eqref{eq_maximum_sep_popIII}.
Meanwhile, they do not find binaries with the small separations of $\sim$~AU owing to the limited spatial resolutions. 
The sink particle method is normally used, and it prevents one from resolving binaries smaller than the sink size.
Even at some larger scales, however, the binary separation is not accurately evaluated. A sink particle could wrongly reduce the gas density in its neighborhood, which results in artificial suppression of the clump migration (e.g., see appendix B). As a result, such simulations may tend to overproduce binaries whose separations are slightly larger than the sink size.

%-----------------------------------------------------------------------------------------------%
If massive Pop~III binaries with $\sim 10$--$10^{3}~M_{\odot}$ finally form, such systems can evolve into equally massive BH binaries \citep[e.g.,][]{HegerWoosley2002}.
However, the minimum separation $a_\text{min} \sim 1$~AU given by eq.~\eqref{eq_minimum_sep_popIII} is still too large to cause the merger, simply because it takes very long time
\begin{align}
t_\text{GW, merge} = 1.25 \times 10^{12}~\mathrm{yr} \left ( \frac{a}{1~\mathrm{AU}} \right )^{4}
\left ( \frac{M_\text{BH}}{100~M_{\odot}} \right )^{-3}
\end{align}
\citep{Peters1964}. To make it possible, several mechanisms to decrease the separation have been proposed, including gas dragging \citep[e.g.,][]{Ostriker1999, TanakaHaiman2009},
three-body scattering \citep[e.g.,][]{Stacy+2016,Tagawa+2016}, and
secular Kozai mechanism \citep[e.g.,][]{Kozai1962,Lidov1962}.
To assess how efficiently the above mechanisms work,
we should investigate not only the binary separation, but also the initial mass of the secondary star 
\citep[e.g.,][]{Hanawa+2010,Satsuka+2017}. 
The surrounding environment is also important for the tight binary formation. If young binaries are embedded in a dense stellar cluster, the efficient three body scattering with the cluster member stars may efficiently work to reduce the binary separations. Such dense clusters are expected to form in slightly metal-enriched halos in the early universe
\citep{Katz+2015, Kashiyama&Inayoshi2016, Sakurai+2017, Reinoso+2018, Boekholt+2018}.

%---------------------------------------------------------------------------------%

\subsubsection{Present-day low-mass star formation}

The disk fragmentation is also believed to occur in the present-day star formation \citep[e.g.,][]{Tohline2002,Kratter+2016}. Recent observations have revealed a number of young stellar objects surrounded by massive gas disks \citep[e.g.,][]{Jorgensen+2009,Eisner2012}.
Some of them show signatures of the disk fragmentation, e.g., sub-structure such as spiral arms \citep{Perez+2016,Tomida+2017} and multiple clumps \citep{Tobin+2016}.

%--------------------------------------------------------------------%

The present-day star formation starts with the collapse of a molecular cloud core.
The collapse advances almost isothermally at $T \sim 10$~K with efficient radiative cooling via dust thermal emission. It continues until a pressure-supported core appears at $n \gtrsim 10^{11}~\mathrm{cm^{-3}}$ \citep[the so-called first core; e.g.,][]{Masunaga+1998}. The gas temperature increases adiabatically with $\gamma = 7/5$ above this critical density.
Applying eqs.~\eqref{eq_a_max} and \eqref{eq_a_min} assuming such an EOS, we obtain the maximum and minimum separations as,
\begin{align}
a_\text{min}  &= 20~\mathrm{AU}   \nonumber \\
&\;\;  \left( \frac{M_{*}}{0.1~M_{\odot}} \right )^{1/9} 
\left ( \frac{c_\text{s}}{0.3~\mathrm{km~s^{-1}}} \right )^{2/3} \left ( \frac{n_{0}}{10^{11}~\mathrm{cm^{-3}}} \right )^{-4/9} , \label{eq_minimum_sep} \\
a_\text{max} &= 360~\mathrm{AU}    \left( \frac{M_{*}}{0.1~M_{\odot}} \right ) \left ( \frac{c_\text{s}}{0.3~\mathrm{km~s^{-1}}} \right )^{-2}. \label{eq_a_max_pd}
\end{align}
These values are consistent with previous numerical simulations, 
which follow the disk fragmentation and subsequent orbital evolution of fragments \citep[e.g.,][]{Vorobyov+2010,Zhu+2012,Tsukamoto+2013,Hall+2017,Fletcher+2019}.

%-----------------------------------------------------------------------%

Our numerical experiments suggest that, if a fragment survives without merging away with the central star, the separation quickly converges to $a_\text{max}$ after the outward migration. Eq.~\eqref{eq_a_max_pd} predicts that $a_\text{max}$ increases with increasing the central stellar mass. This trend is consistent with the recent observations; the typical binary separation gets larger with the higher stellar mass of the binary members \citep{Dunchene+2013}.

%---------------------------------------------------------------------------------------%

Recent observations reveal a large population of binaries whose separation is much smaller than $\sim 10$~AU \citep[e.g.,][]{Sana+2012,Kobulnicky+2014,Eker+2014}.
Our estimation does not exclude the existence of such close binaries.
Eq.~\eqref{eq_scaling}, which assumes the adiabatic EOS, gives the clump radius $R_\text{clump}$ as
\begin{align}
R_\text{clump} &= 10~\mathrm{AU} \nonumber \\
&\left ( \frac{T_0}{10~\mathrm{K}} \right ) \left ( \frac{n_{0}}{10^{11}~\mathrm{cm^{-3}}} \right ) ^{-2/3}
\left ( \frac{M_\text{clump}}{0.002~M_\odot} \right )^{-1/3},
\end{align}
where $M_\text{clump}$ is the clump mass. 
However, the assumption of such a stiff EOS is not always satisfied for the present-day case. 
According to \citet{Zhu+2012}, the KH and orbital timescales of the clump are respectively written as
\begin{align} \label{eq_tcool_pd_clump}
t_\text{KH} &= 122~\mathrm{yr} \left ( \frac{R_\text{clump}}{10~\mathrm{AU}} \right )^{-1.8}, \\
t_\text{orb}  &= 126~\mathrm{yr} \left ( \frac{a}{40~\mathrm{AU}} \right )^{3/2} \left ( \frac{M_*}{0.1~M_\odot} \right )^{-1/2}.
\end{align}
Since these are comparable, the migrating clumps have diverse fates depending on their initial separations \citep[e.g.,][]{Zhu+2012}.
If $a \gg 40~$AU, the clump should have enough time to contract during the migration. For such a case, the clump size is further reduced from the estimate given by eq.~\eqref{eq_scaling} owing to the radiative cooling. 
The clump can get closer to the central star than $\sim 10$~AU, avoiding the RLOF-migration.

%%%%%%%%%%%%%%%%%%%%%%%%
%%%%%%%%%%%%%%%%%%%%%%%%
\section{Summary}  
\label{sec_conclusion}
%%%%%%%%%%%%%%%%%%%%%%%%
%%%%%%%%%%%%%%%%%%%%%%%%

Although the disk fragmentation is a possible process to yield Pop III stellar binary systems, simulation results are often too complex to extract key physical processes. In order to improve such a situation, we have performed a suite of numerical experiments in a well organized manner. 
We have artificially inserted a point particle in a disk, and then followed its orbital evolution. 
The particle migrates through the disk, accreting the gas to develop a surrounding envelope. Such migrating ``clumps'' show the diverse orbital evolution, depending on their initial masses and positions; some migrate inward to merge away with the central star, and the others survive for a while in binaries. 
We have shown that the numerical results are well interpreted with the analytical modeling of the key physical processes. Our findings are summarized as follows:
%----------------------------------------------------------------------------------------------------------------------------%
\begin{enumerate}
\item 
In general, the inserted particles first migrate inward regardless of the initial masses and positions. The migration continues until the clumps merge away with the central star for some cases. For the other cases, however, the inward migration ceases when the gap structure is cleared in the disk. Although the results show the divergent evolution, there is a certain trend that a clump has a larger chance of the survival with the higher initial mass. Apparently the dependence on the initial positions is not straightforward.
%------------------------------------------------------------------%
\item 
The numerical results are well interpreted postulating that the inward migration is driven by two distinct processes. 
One is the Type-I migration: the angular momentum transfer due to the interaction with the disk gas. It works when the Hill radius is smaller than the disk scale height, which is satisfied in an outer part of the disk.
The other process is the RLOF: the tidal disruption of the outer envelope of a clump, with which the orbital angular momentum is carried away with the outflowing gas. It operates when the clump radius is larger than the Hill radius. Such a condition is satisfied in an inner part of the disk, where the tidal force from the central star becomes strong.
%------------------------------------------------------------------%
\item 
Comparisons between the numerical results and analytic evaluations of the above two processes show excellent agreements. For that purpose, we have made use of the $q_\text{env}$--$a$ diagram (Fig.~\ref{fig_rq_evolution}), where $q_\text{env}$ is the mass ratio of the envelope to the clump and $a$ is the separation from the central star. 
The figure predicts that the inward migration ceases in a middle part of the disk, if the clump mass is larger than a threshold value. 
Our simulation results well agree with such analytical evaluations.
%--------------------------------------------------------------------%
\item 
For the cases where the early migration stalls, the clumps turn to migrate outward afterwards. The outward migration continues for a while, but finally ceases at some point. Interestingly, the final separations are independent of the initial masses and positions. We propose that the outward migration occurs as the clump accretes the disk gas which has the high specific angular momentum.  
The larger amount of the mass accretion is necessary for migrating to the outer part of the disk. However, the disk gas available for the accretion is limited by the isolation mass given by eq.~\eqref{eq_Miso}.
Such a balance sets the maximum separation, beyond which the outward migration is not possible. This explains well our numerical results. 
We suggest that the separation of a newly-born stellar binaries should sensitively depend on mass accretion histories. 
%-------------------------------------------------------------------%
\item
Although a specific EOS supposing the DC has been assumed in the current work, our
results can be rescaled to be applied to other cases such as the normal Pop III star formation and even present-day star formation.  It appears that our framework is also consistent with numerical results in such other contexts. 
\end{enumerate}
%-------------------------------------------------------------------%

We thank K. Sugimura, K. Omukai, H. Susa, S. Inutsuka, N. Kanagawa, and H. Tanaka for fruitful discussions and comments. 
This work is financially supported by
the Grants-in-Aid for Basic Research by the Ministry of Education, Science and Culture of Japan 
(17H01102: S.C., 16H05996, 17H06360: T.H.). Numerical computations are carried out on XC50 at the Center for Computational Astrophysics (CfCA) of the National Astronomical Observatory of Japan. 
We use the SPH visualization tool SPLASH \citep{SPLASH} in Figs. \ref{fig_initial_condition}, \ref{fig_snapshots}, and \ref{fig_map_phases}.

\bibliography{biblio2}

\newpage
\appendix

\section{The effect of Disk mass on the final separation} \label{sec_revo_sd}
\begin{figure}
	\centering
		\includegraphics[width=8.7cm]{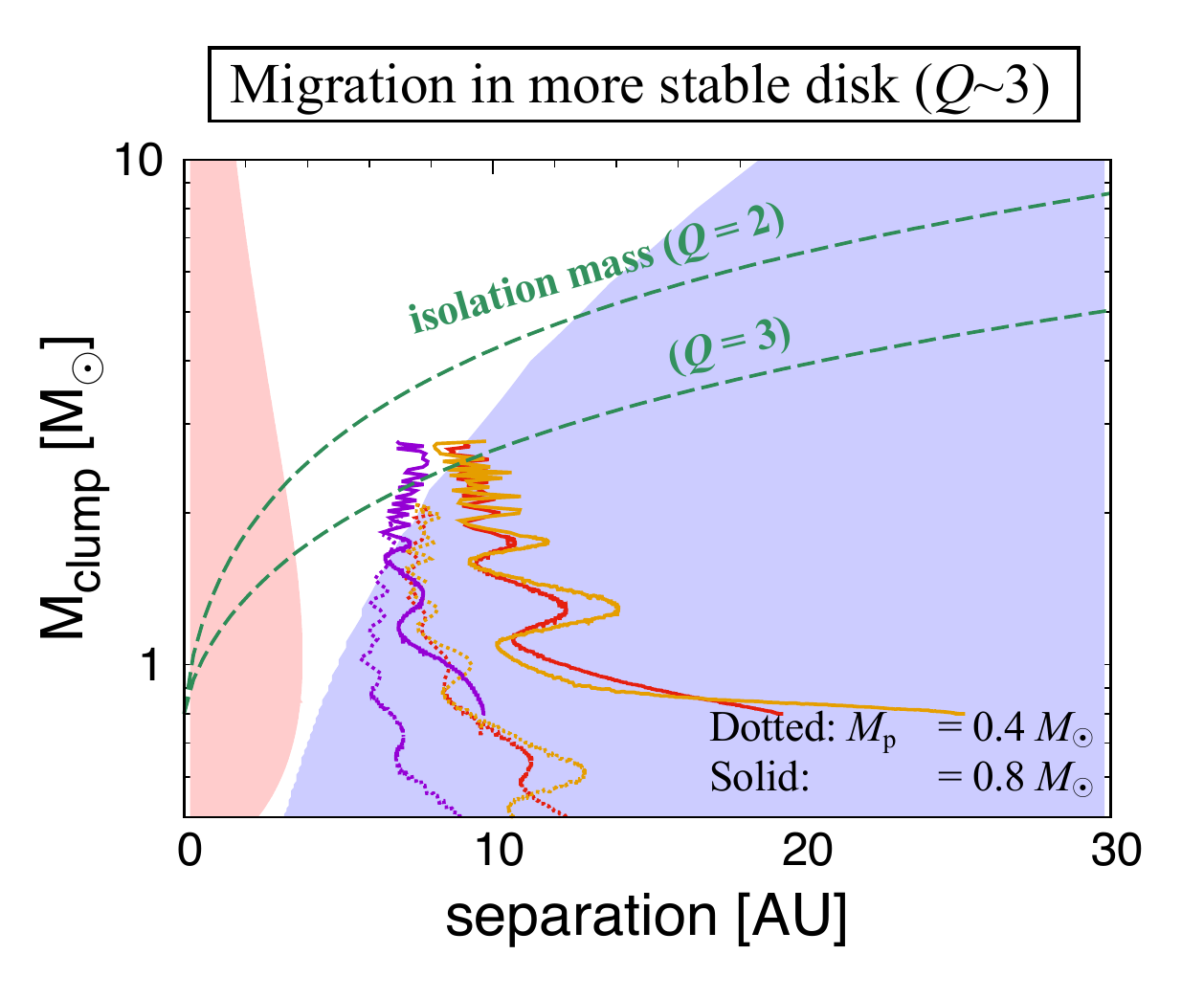}
		\caption{The same as Fig.~\ref{fig_rm_plane}~(b), but for the cases with the less massive disk. The dashed lines represent the isolation masses given by assuming the constant Toomre parameters of $Q=2$ and $3$. The solid and dashed trajectories show the orbital evolution of the migrating clumps for the different initial core masses $M_\text{p}=0.4$ and $0.8~M_{\odot}$, respectively.}
		\label{fig_rm_sd_plane}
\end{figure}

In Section~\ref{sec_final_separation}, we have shown that the outward migration should continue until the binary separation reaches the maximum separation $a_\text{max}$, set by the isolation mass $M_\text{iso}$.
It follows that such an outward motion hardly occurs for cases where $a_\text{max}$ is small because of the low isolation mass. We here examine such cases by performing additional numerical experiments.

%-----------------------------------------------------------------------%

To this end, we modify the set-up process described in Section~\ref{ssec_setup}. Specifically, we follow a longer-term evolution of the late accretion stage for $700$~years.
The masses of the central star and the surrounding disk are $13$ and $3~M_{\odot}$ at this epoch, respectively. Note that the gas supply from the cloud envelope to the disk-star system almost ceases by this time, because of the limited size of the initial cloud (Section~\ref{sssec_collapse}). 
The disk is thus less massive and more stable against the gravitational instability than the default cases considered in the main part. 
The Toomre-Q parameter takes $\simeq 3$ on average throughout the disk. 
We then insert a point particle with $M_p = 0.4$ or $0.8~M_\odot$
into the disk to follow their orbital evolution. The initial positions are varied in the range of $10$--$30~$AU apart from the central star.

%-------------------------------------------------------------------%

Fig.~\ref{fig_rm_sd_plane} shows the clump mass and orbital evolution for such cases in the same style as in Fig.~\ref{fig_rm_plane}. 
We see that for all the cases the early inward migration ceases at the separations of $a \simeq 5-10$~AU. 
Such evolution perfectly agrees with the analytic evaluation, as the trajectories reach the boundary of $R_\text{Hill} = H_\text{d}$ in the figure. The inward migration ceases with the gap opening in the end.
We see there is no overlap between the blue and red area in the figure, explaining why the RLOF-driven migration never operates for the clumps to merge with the central star. 

%----------------------------------------------------------------%

Fig.~\ref{fig_rm_sd_plane} also shows that the binary separations hardly increase after the inward migration ceases. 
The separations only increase by at most $20\%$ among the current cases.
Recall that the separations increase by factors of $4$--$6$ with the marginally stable self-gravitating disk with $Q \simeq 1$ (see Fig.~\ref{fig_sep_long}). Fig.~\ref{fig_rm_sd_plane} well explains such a difference, showing the line of the isolation mass assuming $Q = 3$ limits the further outward migration.
The above results suggest that the outward migration is only expected with a massive gas disk, or in an early evolutionary stage of the protostellar accretion where the gas supply from the cloud envelope still continues. 
This is also consistent with previous studies on the planetary migration, for which a gravitationally stable disk is normally assumed \citep[e.g.,][]{Stamatellos+2018}.
The similar outward migration has been reported but looks modest in comparison to our fiducial cases presented in the main part.

%%%%%%%%%%%%%%%%%%%%%%%%%%%%%%%%%%%%%%%%%%%%%%%%%%%%%%%%%%%%%%
\section{The binary evolution after the mass accretion ends} 
\label{sec_revo_long}
%%%%%%%%%%%%%%%%%%%%%%%%%%%%%%%%%%%%%%%%%%%%%%%%%%%%%%%%%%%%%%

\begin{figure}
	\centering
		\includegraphics[width=9cm]{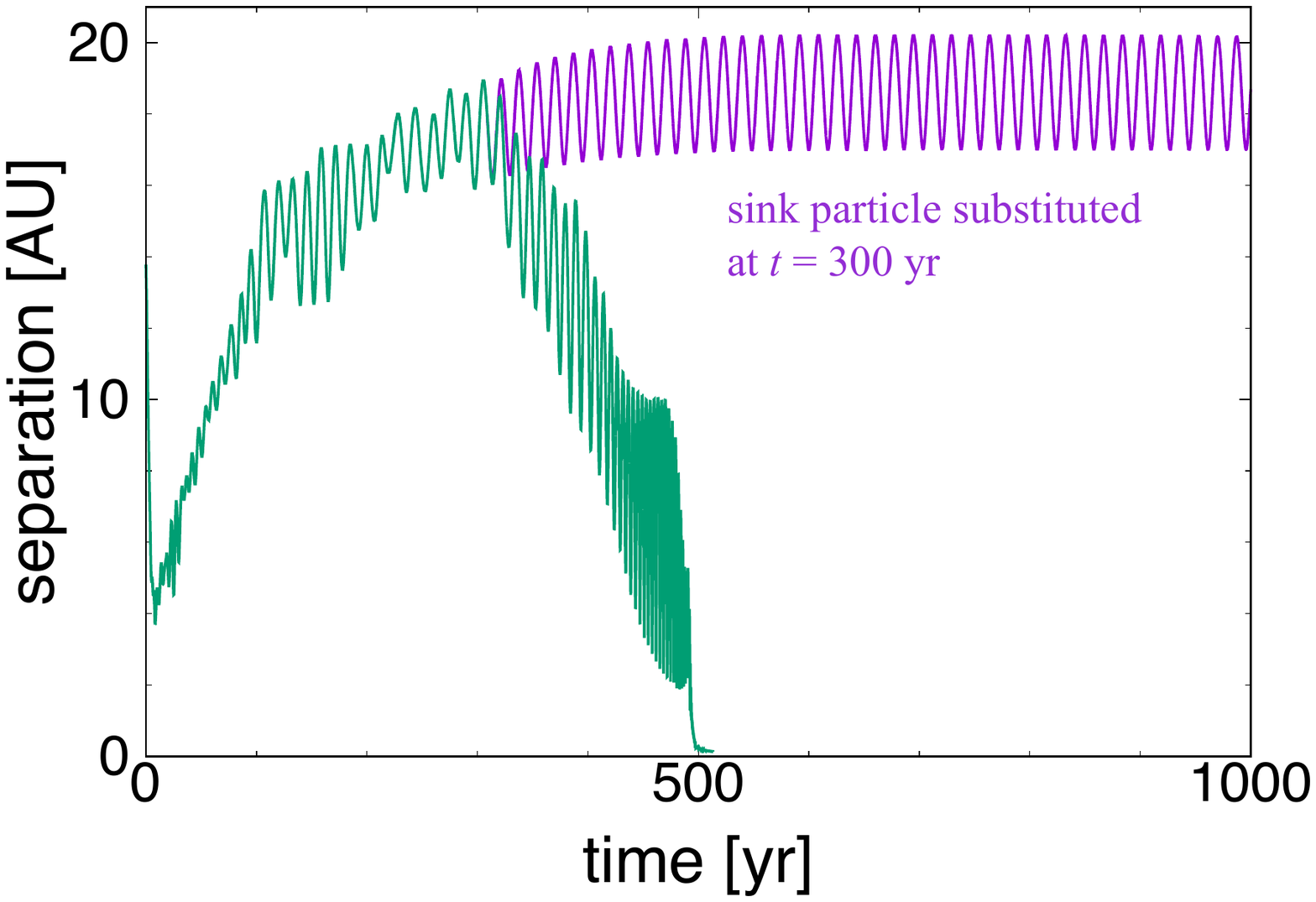}
		\caption{
		Effects of replacing the clump core with a sink particle for the long-term evolution after the mass accretion ceases. The green line represents the default case M02A, where the 
		clump core is modeled by a point particle which does not swallow the gas.
		We can see the separation starts to decrease after $t \simeq 300~$years,
		at which the mass accretion from the envelope onto the binary system ends.
		The purple line represents the case where the core of the clump is replaced by a sink at the epoch of $t=300~$years.
		}
		\label{fig_Revo_w_sink}
\end{figure}

As stressed in Section~\ref{ssec_particle}, we have only focused on the evolution in an earliest phase of the protostellar accretion.  
This is because of our artificial set-up of the simulations, i.e., the initial cloud with $\simeq 50~M_\odot$, while more massive ones with $\sim 10^5~M_\odot$ are normally supposed for the realistic DC model. Accordingly, the mass accretion from the envelope thus ceases after a few $\times$ 100 years in our simulations.
Although it is purely artificial, we here dare to extend our simulations after that. It allows us to investigate impacts of the mass accretion on the orbital evolution of binary systems. 

%-------------------------------------------------------------------------------------------%

We here focus on case M02A, where the early inward migration is reversed in the initial $\sim 10$ years. The outward migration finally ceases at the separation of $\simeq 16$~AU around $t \simeq 200$~years, as described in Section~\ref{sec_final_separation}. Fig.~\ref{fig_Revo_w_sink} shows the subsequent evolution after that. The green line represents the default case, where the binary separation turns to decrease for $t \gtrsim 300~$years. The clump migrates inward again, and then merges with the central star at $t \simeq 500~$years. Although not presented, we have confirmed that the similar evolution occurs for the other ``survival'' models (M02C, M02D, M02E, M01C, and M01D). 

%---------------------------------------------------------------------------------------------%

We infer that this phenomenon is caused by weak RLOF. It apparently contradicts with our argument in Section~\ref{sec_Rochelobe}, saying that the RLOF-driven inward migration should occur in an inner part of the disk. In fact, the necessary condition $R_\text{clump} > R_\text{Hill}$ is only satisfied for $a < 10$~AU for case M02A. However, the clump is actually surrounded by a small Keplerian disk, which extends until the Hill radius  \citep[e.g.][]{Ryan+2017}. A small amount of the gas escapes from the edge of such a ``mini-disk'', leading to the weak RLOF-driven migration. 
Such a weak effect should not appear under the mass accretion, which drives the more prominent outward migration. 

%-----------------------------------------------------------------------------------%

To prove the above hypothesis, we perform an additional artificial experiment. 
We replace a clump core with a sink particle at the epoch of $t = 300$~years, and follow the subsequent orbital evolution. The sink particle rapidly swallows the surrounding gas including the mini-disk. Indeed, Fig.~\ref{fig_Revo_w_sink} shows that the inward migration no longer occurs and the binary separation stays around $20~$AU until the epoch of $t=1000$~years. 

%-------------------------------------------------------------------------------------%

If the mass accretion ceases before the clump contracts,
the clump gradually migrates inward and finally merges away as shown above.
That is, in order to keep the stable binary orbit,
the KH time should be smaller than the accretion timescale $t_\text{acc}$.
For the DC case, the mass accretion continues for $10^5$--$10^6~$years \citep[e.g.,][]{Chon+2018} and the KH time is,
\begin{align}
t_\text{KH} &\simeq 1\times 10^{3} ~\mathrm{yr}  \left ( \frac{M_\text{clump}}{10~M_{\odot}} \right )^{-7/2}
\left ( \frac{R_\text{clump}}{\mathrm{AU}} \right )^{-1/2}.
\end{align}
This clearly shows that the clump has enough time to contract much before the mass accretion ceases.
Note that the stellar evolution calculations predict that the rapid mass accretion inflates a stellar surface layer even with the timescale balance of $t_\text{KH} \ll t_\text{acc}$ \citep[e.g.,][]{Hosokawa+2012, Hosokawa+2013}.
Still, most of the mass is concentrated at the stellar center and only a small amount of the gas near the surface can contribute to the RLOF.
Thus in this situation, we do not expect such an efficient inward migration observed here.

\newpage

\end{document}